\documentclass[journal]{IEEEtran}
\usepackage{amsmath}
\usepackage{diagbox}
\usepackage{graphicx}
\usepackage{float}
\newfloat{figtab}{htb}{fgtb}
\makeatletter
\newcommand\figcaption{\def\@captype{figure}\caption}
\newcommand\tabcaption{\def\@captype{table}\caption}
\makeatother

\usepackage{amssymb,amsthm}
\usepackage{color}
\usepackage{graphicx}
\usepackage{empheq}
\usepackage[linesnumbered,ruled,lined]{algorithm2e}
\usepackage[outdir=./]{epstopdf}
\usepackage{caption}
\usepackage{lipsum}
\usepackage{cite}
\usepackage{multirow}
\usepackage{subcaption}
\usepackage{fancyhdr}
\usepackage[colorlinks,
linkcolor=blue,       
anchorcolor=blue,  
citecolor=blue,        
]{hyperref}

\newtheorem{theorem}{Theorem}

\newtheorem{lemma}{Lemma}

\SetKwInput{Input}{input}
\SetKwInput{Output}{output}

\usepackage{array}
\newcolumntype{L}[1]{>{\raggedright\let\newline\\\arraybackslash\hspace{0pt}}m{#1}}
\newcolumntype{C}[1]{>{\centering\let\newline\\\arraybackslash\hspace{0pt}}m{#1}}
\newcolumntype{R}[1]{>{\raggedleft\let\newline\\\arraybackslash\hspace{0pt}}m{#1}}

\newtheorem{proposition}{Proposition}
\newtheorem{corollary}{Corollary}

\newtheorem{remark}{Remark}
\IEEEoverridecommandlockouts
\def\specialpapernotice#1{\if@confmode%
	\def\@specialpapernotice{{\sublargesize\textit{#1}\vspace*{1em}}}%
	\else%
	\def\@specialpapernotice{{\\*[1.5ex]\sublargesize\textit{#1}}\vspace*{-2ex}}%
	\fi}
\usepackage{amsmath,amsfonts}
\usepackage{algorithmic}
\usepackage{array}
\usepackage{textcomp}
\usepackage{stfloats}
\usepackage{url}
\usepackage{verbatim}
\usepackage{graphicx}
\usepackage{stfloats}
\usepackage{balance}
\usepackage{booktabs}
\usepackage{etoolbox}
\makeatletter\patchcmd{\@makecaption}{\scshape}{}{}{}
\newcommand{\pushright}[1]{\ifmeasuring@#1\else\omit\hfill$\displaystyle#1$\fi\ignorespaces}
\usepackage{amsthm,amssymb,amsfonts}
\usepackage{bm}
\usepackage{cases}

\begin{document}
\title{Rate Splitting Multiple Access: Optimal Beamforming Structure and Efficient Optimization Algorithms}
\author{Tianyu Fang, \IEEEmembership{Graduate Student Member, IEEE}, and Yijie Mao, \IEEEmembership{Member, IEEE}
\thanks{This work has been supported in part by the National Nature Science Foundation of China under Grant 62201347; and in part by Shanghai Sailing Program under Grant 22YF1428400.
	\par This work was presented in part at the 2024 IEEE international Conference on Acoustics, Speech, and Signal Processing (ICASSP) \cite{fang2024optimal}.
\par T. Fang and Y. Mao are with the School of Information Science and Technology, ShanghaiTech University, Shanghai 201210, China (e-mail: fangty@alumni.shanghaitech.edu.cn, maoyj@shanghaitech.edu.cn).}}


\maketitle

\begin{abstract}
Joint optimization for common rate allocation and beamforming design have been widely studied in rate splitting multiple access (RSMA) empowered multiuser multi-antenna transmission networks. 
Due to the highly coupled optimization variables and non-convexity of the joint optimization problems, emerging algorithms such as weighted minimum mean square error (WMMSE) and successive convex approximation (SCA) have been applied to RSMA which typically approximate the original problem with a sequence of disciplined convex subproblems and solve each subproblem by an optimization toolbox. While these approaches are capable of finding a viable solution, they are unable to offer a comprehensive understanding of the solution structure and are burdened by high computational complexity.
In this work, for the first time, we identify the optimal beamforming structure and common rate allocation for the weighted sum-rate (WSR) maximization problem of RSMA. We then propose a computationally efficient optimization algorithm that jointly optimizes the beamforming and common rate allocation without relying on any toolbox.
Specifically, we first approximate the original WSR maximization problem with a sequence of convex subproblems based on fractional programming (FP).  By exploiting the Karush-Kuhn-Tucker (KKT) conditions of each subproblem, the optimal beamforming structure is derived. An efficient hyperplane fixed point iteration method is then proposed to find the optimal Lagrangian dual variables. 
Numerical results show that the proposed algorithm achieves the same performance but takes only 0.5\% or less simulation time compared with the state-of-the-art WMMSE, SCA, and FP algorithms.
The proposed algorithms pave the way for the practical and efficient optimization algorithm design for RSMA and its applications in 6G. 
\end{abstract}

\begin{IEEEkeywords}
 Rate-splitting multiple access, beamforming optimization, weighted sum-rate maximization, optimal beamforming structure. 
\end{IEEEkeywords}

\section{Introduction}\label{Sec:intro}
\par In recent years, there has been a significant interest in rate splitting multiple access (RSMA) as an influential interference management and non-orthogonal transmission technique for the sixth generation (6G) wireless networks \cite{bruno2022tutorial,Dizdar2020}. By splitting user messages into common and private parts at the transmitter and allowing receivers to decode the common streams of interfering users as well as the intended private streams, RSMA allows for partially decoding the interference and partially treating the remaining interference as noise \cite{Clerckx2016}. Numerous studies have demonstrated that RSMA achieves the same or superior performance compared to linearly precoded space division multiple access (SDMA), power-domain non-orthogonal multiple access (NOMA), and orthogonal multiple access (OMA) across various performance metrics. These metrics include but not limited to spectral efficiency, energy efficiency (EE), max-min fairness, and transmission robustness across a broad spectrum of network loads and user deployments, encompassing both underloaded and overloaded scenarios. Notably, the performance advantage of RSMA becomes particularly evident in overloaded regions or when users encounter significant discrepancies in channel strength \cite{Mao2022}.


\par One major driver to attain all the aforementioned performance gains of RSMA is an exquisite joint optimization algorithm for the beamforming design and common rate allocation, which has been widely studied in the state-of-the-art works \cite{Matth2022Globally, Wang2023,Hamdi2016,mao2018,fang2022,Mao2019uni-multicast,Mao2020,Li2020,Fu2020,Dizdar2020a,Onur2021Radar,Xu2021,CernaLoli2021}. 
Due to the inter-user interference, optimization problems of RSMA are typically non-convex, leading to the development of global optimal and suboptimal algorithms.
In \cite{Matth2022Globally}, a globally optimal optimization algorithm using successive incumbent transcending (SIT) and branch and bound (BB) was proposed for maximizing the weighted sum-rate (WSR) and EE of RSMA for a two-user multi-antenna broadcast channel (BC).
Another optimal beamforming algorithm based on monotonic optimization was proposed in \cite{Wang2023}, specifically tailored for RSMA with finite blocklength. Although these global optimal algorithms exhibit appealing performance, they are afflicted by notable computational complexity, making them unfavourable for real-world applications.
The state-of-the-art works therefore mainly focus on designing suboptimal algorithms to achieve nearly optimal performance. 
Algorithms based on weighted minimum mean square error (WMMSE) \cite{Hamdi2016,mao2018,fang2022,Mao2019uni-multicast}, successive convex approximation (SCA) \cite{Mao2019uni-multicast,Mao2020,Li2020}, semidefinite relaxation (SDR)\cite{Fu2020,Dizdar2020a}, and  alternating direction method of multipliers (ADMM) \cite{Onur2021Radar,Xu2021,CernaLoli2021} have already been proposed to solve different optimization problems for RSMA.
These algorithms all adopt the approach of approximating the original non-convex optimization problem to one or a sequence of convex problems, which can be solved using standard convex optimization methods usually implemented by a certain solver in the CVX toolbox \cite{grant2014cvx}.
It has been shown in \cite{Matth2022Globally} that the solutions of suboptimal algorithms based on first-order approximations, i.e., WMMSE and SCA, are sufficiently close to the global
optimal solutions of WSR and EE maximization problems for RSMA.
However, the computational complexity of these algorithms remains undesirable in practical applications due to the iterative use of CVX optimization solvers.
\par 
A number of existing works, therefore, switch to the closed-form and low-complexity beamforming design. Heuristic algorithms including singular value decomposition (SVD) \cite{Hamdi2016}, random beamforming (RBF) \cite{Hao2015,Dizdar2021,Lu2018,Piovano2016}, and matched beamforming (MBF) \cite{Lu2018,Clerckx2020,Dai2016,Papazafeiropoulos2018} were proposed to design the precoder of the common stream for RSMA. Classical linear precoding approaches such as zero-forcing (ZF) \cite{Hamdi2016,Dizdar2021}, and minimum mean square error (MMSE) \cite{Lu2018} were used for the precoders of the private streams. However, these low-complexity algorithms inevitably bring performance loss compared with the  global optimal or suboptimal algorithms.
To maintain a relatively low computational complexity while not hampering the near optimal performance of RSMA, an optimization algorithm without using toolbox, namely generalized power iteration (GPI) approach, was proposed in \cite{Park2023} to maximize the sum-rate.
Such approach first approximates the non-smooth common rate of RSMA by a LogSumExp technique and reformulate the sum-rate maximization problem as a tractable form of the log-sum of Rayleigh quotients. The first-order optimality condition of the transformed problem is then established, and the iterative algorithm GPI is proposed to compute the numerical solution that meet such condition iteratively.
However, such iterative algorithm cannot guarantee either the convergence or the Karush-Kuhn-Tucker (KKT) optimality of the original sum-rate optimization problem due to the lower-bound approximation of the objective function as well as the tradeoff between the approximation accuracy and the approximation coefficient for convergence.

\par A straightforward way to design an algorithm with low computational complexity while maintaining near-optimal performance is to find and utilize the optimal beamforming structure of the original optimization problem. The optimal unicast multiuser beamforming structure for SDMA has been identified in \cite{Bjornson2014} for two different optimization problems, namely, the transmit power minimization problem subject to the signal-to-interference-plus-noise ratio (SINR) constraint of each user and the sum-rate maximization problem subject to the transmit power constraint. It has been shown in \cite{Bjornson2014} that the optimal beamforming structures for the aforementioned two optimization problems are equivalent. Recently, the authors in \cite{Dong2020} identified the optimal beamforming structure for a multigroup multicasting network based on linearly precoded SDMA, with a special focus on the transmit power minimization problem. It has been shown in \cite{Dong2020} that the optimal multi-group multicast beamforming can be viewed as a generalized version of the optimal unicast beamforming, which is a weighted MMSE filter with a similar covariance matrix structure, except that the beamforming direction is the channel direction of a multicast group rather than an individual user. However, the aforementioned optimal beamforming structures for SDMA cannot be directly applied to characterize the optimal beamforming structure of RSMA due to the introduced common streams. Additionally, the optimal common rate allocation of RSMA remains unknown in the WSR problem. To the best of the authors' knowledge, the optimal solution structure for the joint beamforming design and rate allocation optimization problem in RSMA, remains an open problem.

This paper aims to fill this gap and also design an efficient beamforming and common rate allocation algorithm to achieve near optimal WSR of RSMA while significantly reducing the computational complexity for RSMA. To achieve this goal, in this work, we focus on the WSR maximization problem of 1-layer RSMA for multi-antenna BC, and the key contributions of this paper are outlined as follows:
\begin{itemize}
	\item By leveraging the fractional programming (FP) approach and Lagrange duality, we identify the optimal beamforming structure and common rate allocation for the WSR problem of 1-layer RSMA. To the best of our knowledge, this is the first paper that characterizes the optimal beamforming structure and common rate allocation for RSMA.
	\item By exploiting the optimal solution structure of RSMA, we develop an efficient CVX-free algorithm based on fixed point iteration to determine the Lagrangian dual variables so as to find the suboptimal beamforming and common rate allocation. We further provide the convergence proof of the proposed algorithm and demonstrate its ability to reach at least one stationary point of the original WSR problem of RSMA.
	\item Numerical simulations are carried out to show that the proposed algorithm achieves the same or even better performance than existing optimization algorithms including the aforementioned WMMSE, FP, SCA, GPI, while dramatically reducing the computation time. The proposed algorithm only takes 0.5\% or less simulation time compared with the state-of-the-art algorithms.
\end{itemize}

\par \textit{Organization:} The rest of this paper is organized as follows. Section \ref{Sec:system model} presents the system model of the WSR optimization problem of 1-layer RSMA. In Section \ref{Sec:optimal}, we derive and characterize the main results of the optimal beamforming structure and common rate allocation. Section \ref{Sec:numerical} introduces the proposed hyperplane fixed point iteration algorithm to determine the involved Lagrangian dual variables in the optimal beamforming structure. The proposed algorithm is extended to imperfect channel state information at the transmitter (CSIT) scenarios in Section \ref{Sec:imperfect}.  Simulation results are presented in Section \ref{Sec:simulation}, and finally, in Section \ref{Sec:concu}, we conclude the paper and discuss the future work.

\par \textit{Notations:} Scalars and functions are written in normal font while vectors and matrices are denoted by bolded lower-case and upper-case letters, respectively. $\mathbb{C}$ represents the complex-value space, and $ \mathcal{A} $ denotes a predefined set. $ \mathbb{E}[s] $ denotes expectation over random variable $ s $. We use $ |x| $ to denote the magnitude of a complex number $ x $ and $ \|\mathbf y\|_{1}  $ to denote the $ l1 $-norm of a complex vector $ \mathbf y $. $ \mathcal{CN}(0,\sigma^2) $ represents the circularly symmetric complex Gaussian distribution (CSCG) with zero mean and variance $ \sigma^2 $. $ (\cdot)^H $ denotes the conjugate transpose. $ (\cdot)^\star $ refers to the optimal solution for a convex subproblem. $ (\cdot)^\diamond$ and $(\cdot)^\circ $ respectively represent the local and global optimal solution for the original non-convex problem. $ \mathrm{diag}\{ \mathbf y \} $ is a diagonal matrix with the entries of $ \mathbf y $ along the main diagonal.     

\section{System  Model and Problem Formulation}\label{Sec:system model}
\label{system}
\par In this work, we consider the most popular RSMA strategy, namely, 1-layer RSMA \cite{Clerckx2016} for a multiple-input single-output (MISO) BC consisting of one base station (BS) and $K$ users. The BS has $ L $ antennas and each user has a single-antenna. The users are indexed by $ \mathcal{K}=\{1,2,\ldots,K\} $. 
  At the BS, message $ W_k $ for user $ k $ is split into a common part $ W_{c,k} $ and a private part $ W_{p,k} $. The common parts of all users are combined and jointly encoded into a common stream $ s_0 $ while the private parts are independently encoded into the private streams $ s_1,\cdots,s_K $. All streams $ \{s_0,s_1,\cdots, s_K\} $ are assumed to be independent, and each stream satisfies $\mathbb{E}[s_ks_k^H]=1$, and $ \mathbb{E}[s_ks_i^H]=0, \forall k\neq i$ and $k,i\in\mathcal{K}\cup\{0\}$. Denote the corresponding beamforming vector of $ s_k $ as $ \mathbf w_k\in\mathbb{C}^{L} $  and the beamforming matrix for all streams as $ \mathbf W=[\mathbf w_0,\mathbf w_1,\cdots,\mathbf w_K] \in \mathbb{C}^{L\times (K+1)} $. The signal transmitted by the BS is $ \mathbf{x}=\sum_{k=0}^{K}\mathbf{w}_ks_k $, and it satisfies the transmit power constraint, i.e., $ \mathrm{tr}(\mathbf {WW}^H)\leq P_t $. $ P_t $ represents the upper limit of the transmit power at the BS.

\par The downlink channel vector from the BS to user $ k $ is denoted as $ \mathbf{h}_k^H\in\mathbb{C}^{1\times L} $, whose entries are drawn from $ \mathcal{CN}(0,1) $. As the main objective of this work is to investigate the optimal beamforming structure and efficient algorithms, we assume that the CSIT is perfect and invariant during each transmission block. The extension to imperfect CSIT will be specified in Section \ref{Sec:imperfect} later on. The signal received at user $k$ is expressed as $ l_k=\mathbf h_k^H\sum_{i=0}^{K}\mathbf w_is_i+n_k,  k\in\mathcal{K} $, where $ n_k\sim \mathcal{CN}(0,\sigma_k^2) $ denotes the additive white Gaussian noise (AWGN) with $ \sigma_k^2 $ being the noise power of user $k$.

\par Following the principle of 1-layer RSMA, each user first decodes the common stream $ s_0 $, and employs one layer of successive interference cancellation (SIC) to remove the common stream from the received signal, and then decodes the intend private stream $ s_k $. The SINRs of the common and private streams at user $ k $ are given as 
\begin{equation}
	\label{eq:SINRs}
		\gamma_{i,k}=\frac{|\mathbf h_k^H\mathbf w_i|^2}{\sum_{j=1,j\neq i}^K|\mathbf h_k^H\mathbf w_j|^2+\sigma_k^2},\,\,\, k\in\mathcal K,i\in\{0,k\},
\end{equation}
where the subscript $\{j=1,j\neq i\}$ reduces to  $\{j=1\}$ when $i=0$ or reduces to $ \{j=1,j\neq k\} $ when $ i=k, k\in\mathcal K $. The corresponding instantaneous achievable rates of common and private streams, also known as common and private rates, are given as $ r_{i,k}=\log\left(1+\gamma_{i,k}   \right),i\in\{0,k\}, k\in\mathcal{K}$. 
To ensure that all users successfully decode $ s_0 $, the common rate should be less than or equal to $\min_{k\in \mathcal{K}} r_{0,k} $. We denote $ c_k $ as the portion of common rate allocated to user $ k $ and the sum of these portions must be less than or equal to the common rate, i.e., $ 	\sum_{k=1}^Kc_k\leq\min_{k\in \mathcal{K}} r_{0,k} $. By synthesizing the decoded $ \widehat{W}_{c,k} $ and $ \widehat{W}_{p,k} $, user $ k $ reconstructs $ \widehat{W}_k $ \cite{Clerckx2016}. Thus, the total transmission rate of user $ k $ is the sum of private rate from decoding private message and the rate allocated from decoding common message, which is given by $ 	r_k^{\text{tot}}=c_k+r_{k,k} $.

\par Let $\delta_k  $ denote to the weight of user $ k $, the weighted sum-rate maximization problem for 1-layer RSMA is formulated as \cite{Hamdi2016}:
\begin{subequations}
	\label{P1}
	\begin{align}
		\max\limits_{\mathbf W,\mathbf c}\,\, &\sum\limits_{k=1}^K \delta_k(c_k+r_{k,k})\\
		\text{s.t.}\,\,		\label{leq: allocation constraint} &c_k\geq 0,\,\,\, \forall k\in \mathcal{K},\\
		\label{leq: common rate}	&\sum_{i=1}^K c_i\leq r_{0,k},\,\,\, \forall k\in\mathcal{K},\\
		\label{eq:power constraint}	& \mathrm{tr}(\mathbf {WW}^H)\leq P_{t},
	\end{align}
\end{subequations}
where $ \mathbf c=[c_1,c_2,\cdots,c_K]^T $.
The aforementioned non-smooth common rate constraint is transformed into an equivalent form in (\ref{leq: common rate}) by introducing $ K $ smooth constraints. Constraint (\ref{leq: allocation constraint}) shows the rate allocated to user $ k $ should be non-negative and (\ref{eq:power constraint}) is the transmit power constraint at the BS.

\section{Optimal Beamforming Structure for RSMA}\label{Sec:optimal}
\label{Section framework}
In this section, We focus on solving the non-convex WSR problem (\ref{P1}). Depart from all existing works that focus on solving problem \eqref{P1} numerically, we take a different approach by utilizing FP and Lagrangian duality to unveil the optimal solution structure of \eqref{P1} as delineated in the following subsections. Specifically, the non-convex problem \eqref{P1} is first transformed to a sequence of convex subproblems based on FP with Lagrangian dual and quadratic transform. We then derive the optimal solution structure based on the KKT optimality of the transformed convex subproblems. Afterwards, the optimality of the beamforming structure is proved and analyzed.

\subsection{Optimal Common Rate Allocation}
\par To find the optimal common rate allocation $ \mathbf c^\circ $, we first visit problem (\ref{P1}) to obtain the following Proposition \ref{pro3}.
\begin{proposition}\label{pro3}
	Assuming all users have different weights, for any fixed $ \{\mathbf W\} $, the optimal WSR for problem \eqref{P1} is always achieved by the following optimal common rate allocation $ \mathbf c^{\circ} $, given as
	\begin{equation}
		\label{opt a}	c_i^{\circ}=\left\{	\begin{aligned}
			&\min_{k\in\mathcal K}\{r_{0,k}\},\,\,\, i = \arg\max_{i\in\mathcal K}\{\delta_i\},\\
			&0,\quad \quad\quad\quad\quad\!\!\!\!\!\!\,\,\,\text{otherwise.}
		\end{aligned}\right.
	\end{equation}
\end{proposition}
\par \textit{Proof:} For any given beamforming matrix $ \mathbf W $, problem (\ref{P1}) is reduced to the following linearly programming (LP):
\begin{subequations}
	\label{P6}
	\vspace{-0.1cm}
	\begin{align}
		\max\limits_{\mathbf c}\,\, &\sum\limits_{k=1}^K \delta_k(c_k+r_{k,k})\\
		\text{s.t.}\,\,&c_k\geq 0,\,\,\ \forall k\in\mathcal{K},\label{P4:C3}\\
		&	\sum_{i=1}^K c_i\leq \min_{k\in\mathcal K}\{r_{0,k}\}. \label{P4:C1}
	\end{align}
\end{subequations}
Here, $ r_{k,k} $ and $ r_{0,k} $ are treated as constants. It is easy to find that
\begin{equation}\label{linear programing}
	\sum_{k=1}^K \delta_k c_k \leq \max_{k\in\mathcal K}\{\delta_k\}\cdot\sum_{k=1}^K c_k\leq \max_{k\in\mathcal K}\{\delta_k\}\cdot\min_{k\in\mathcal K}\{r_{0,k}\}.
\end{equation}
The equalities in \eqref{linear programing} are achieved if and only $\mathbf c=\mathbf c^{\circ}  $, which completes the proof.  $ \hfill\blacksquare $

\par By substituting the optimal common rate allocation back into problem \eqref{P1}, we obtain the following equivalent problem:
\begin{subequations}
	\vspace{-0.3cm}
	\label{P pureW}
	\begin{align}
		\max\limits_{\mathbf W,y}\,\, &\max_{k\in\mathcal{K}} \{\delta_k\} \cdot y+ \sum\limits_{k=1}^K \delta_kr_{k,k}\\
		\text{s.t.}\,\,		
		&y\leq r_{0,k},\,\,\, \forall k\in\mathcal{K},\\
	\label{P Wc2}	& \mathrm{tr}(\mathbf {WW}^H)\leq P_{t},
	\end{align}
\end{subequations}
where $ y $ is a slack variable to represent $ \min_{k\in\mathcal K}\{ r_{0,k}\} $. Problem \eqref{P pureW} is independent of the common rate allocation. We can therefore focus on the beamforming optimization in the following.
\begin{remark}
	Proposition \ref{pro3} indicates that only the user with highest weight will receive the entire common rate when users have different weights. However, if there are users with the same highest weight, any arbitrary distribution of the common rate among those users would yield the same WSR. In the extreme case when all users have the same weights, the corresponding WSR problem reduces to a sum-rate problem which is a special case of \eqref{P pureW} when $ \delta_k=1, \forall k\in \mathcal K $. Therefore, there is no need to delve into different special cases of user weights.
\end{remark} 

\subsection{Fractional Programming Method}\label{FPM}

\par We first apply the FP technique named Lagrangian dual transform proposed in \cite{Shen2018a} to extract the fractional terms from the logarithmic rate expressions. By introducing  auxiliary variable  $ \alpha_{i,k} $  to represent the corresponding SINR $ \gamma_{i,k} $, the  achievable rate $ r_{i,k} $ at user $ k $ for decoding $ s_i, i\in\{0,k\}$ is reformulated as
\begin{equation}
\label{eq:FP1}
	f_{i,k}(\mathbf W,\alpha_{i,k})\triangleq \log(1+\alpha_{i,k})-\alpha_{i,k}+\frac{(1+\alpha_{i,k})\gamma_{i,k}}{1+\gamma_{i,k}},
\end{equation}
where $ f_{i.k}(\mathbf W,\alpha_{i,k}) $ is a lower bound of the original rate $ r_{i,k} $, i.e., $ f_{i,k}(\mathbf W,\alpha_{i,k})\leq r_{i,k}  $ and the equality is achieved if and only if $ \partial f_{i,k}/\partial \alpha_{i,k}=0,k\in\mathcal{K} ,i\in\{0,k\}$. Therefore, $ \alpha_{i,k} $ can be optimally determined by
\begin{equation}
	\label{eq:optalpha}
	\begin{aligned}
		\alpha_{i,k}^{\star}&=\gamma_{i,k}, 
	\end{aligned}
\end{equation}

\par Then, we apply the FP technique named quadratic transform proposed in \cite{Shen2018} to \eqref{eq:FP1}. By introducing auxiliary variable  $ \beta_{i,k} $ to decouple the fractional term in (\ref{eq:FP1}), the recast achievable rate of user $ k $ to decode stream $ s_i, i\in\{0,k\}$ becomes
	\begin{equation}
		\label{eq:FP2}
		\begin{aligned}
			&g_{i,k}(\mathbf W,\alpha_{i,k},\beta_{i,k})\triangleq2\sqrt{1+\alpha_{i,k}}\Re\{\beta_{i,k}^H\mathbf h_k^H\mathbf w_i  \}-\alpha_{i,k}\\
			&\,\,\,\,\,-|\beta_{i,k}|^2\left(\sum_{j\in\mathcal{K}\cup\{i\}}|\mathbf h_k^H\mathbf w_j|^2+\sigma_k^2\right)+\log(1+\alpha_{i,k}),
		\end{aligned}
	\end{equation}
 where the range of summation $\mathcal{K}\cup\{i\}$ reduces to  $\mathcal{K}$ when $i=k, k\in\mathcal{K}$. Following \cite[Lemma 1]{Shen2018}, we could also obtain that $ g_{i,k}(\mathbf W,\alpha_{i,k},\beta_{i,k})  $ is a lower bound of $ f_{i,k}(\mathbf W,\alpha_{i,k}) $ in \eqref{eq:FP2}, i.e. $ 	g_{i,k}(\mathbf W,\alpha_{i,k},\beta_{i,k}) \leq f_{i,k}(\mathbf W,\alpha_{i,k})$ and the equality is attained if and only if $ \partial g_{i,k}/\partial \beta_{i,k}=0,i\in\{0,k\},k\in\mathcal{K} $, yielding the optimal $ \beta_{i,k}^\star $
\begin{equation}
	\label{eq:optbeta}
	\begin{aligned}
		\beta_{i,k}^{\star}&=\frac{\sqrt{1+\alpha_{i,k}}\mathbf{h}_{k}^{H}\mathbf w_i}{\sum_{j\in\mathcal{K}\cup\{i\}}|\mathbf h_k^{H}\mathbf w_j|^2+\sigma_k^2}.
	\end{aligned}
\end{equation}
Substituting $ r_{i,k} $ in (\ref{P1}) with $ g_{i,k}(\mathbf W,\alpha_{i,k},\beta_{i,k})$ leads to the following Proposition \ref{pro: equivalence}.
\par
 \begin{proposition}\label{pro: equivalence}
 	The weighted sum-rate maximization problem (\ref{P1}) is equivalent  to
\begin{subequations}
	\label{P3}
	\vspace{-0.5cm}
	\begin{align}
		\max\limits_{\mathbf W,y,\bm{\alpha}, \bm{\beta}}\,\, &\max_{k\in\mathcal K} \{\delta_k\}\cdot y+\sum\limits_{k=1}^K \delta_kg_{k,k}(\mathbf W,\alpha_{k,k},\beta_{k,k})\\
	\label{P3c1}	\text{s.t.}\,\,	&y\leq g_{0,k}(\mathbf W,\alpha_{0,k},\beta_{0,k}),\,\,\, \forall k\in\mathcal{K},\\
	\label{P3c3}	& \mathrm{tr}(\mathbf {WW}^H)\leq P_{t},
	\end{align}
\end{subequations}
where $g_{i,k}(\mathbf W,\alpha_{i,k},\beta_{i,k}), i\in\{0,k\}, k\in\mathcal{K}$ is defined in (\ref{eq:FP2}), $\bm{\alpha}\triangleq\{\alpha_{i,k}|i\in\{0,k\}, k\in\mathcal{K}\}$ and $\bm{\beta}\triangleq\{\beta_{i,k}|i\in\{0,k\}, k\in\mathcal{K}\}$.
 \end{proposition}
\par 
\textit{Proof:} The equivalence can be readily established by substituting \eqref{eq:optalpha} and \eqref{eq:optbeta} back into problem \eqref{P3}.$ \hfill\blacksquare $ 
\par Problem \eqref{P3} remains non-convex, but it is obvious that, problem \eqref{P3} is block-wise convex when only optimizing one block of variables from $ \{ \bm\alpha\} $,$  \{\bm\beta\} $,$  \{ \mathbf W,y\} $ with the other two blocks being fixed. Therefore, we could adopt a block coordinated descent (BCD) framework to optimize the aforementioned three blocks alternatively. Specifically, for given $ \{ \bm\beta\} $, $ \{\mathbf W,y\} $, the optimal $ \bm\alpha $ can be obtained by \eqref{eq:optalpha}. For given $ \{ \bm\alpha\} $, $ \{\mathbf W,y\} $, the optimal $ \bm \beta $ can be obtain by \eqref{eq:optbeta}. For given $ \{ \bm\alpha,\bm\beta\} $, the optimal $ \{\mathbf W,y\}$ can be obtained by solving the following convex problem:
\begin{subequations}
	\label{P5} 
	\begin{align}
		\max\limits_{\mathbf W,y}\,\, &\max_{k\in\mathcal K} \{\delta_k\}\cdot y+\sum\limits_{k=1}^K \delta_kg_{k,k}(\mathbf W)\\
		\text{s.t.}\,\,&	y\leq g_{0,k}(\mathbf W),\,\,\, \forall k\in\mathcal{K} \label{P5:C1}\\
		& \mathrm{tr}(\mathbf W\mathbf W^H)\leq P_t \label{P5:C2}.
	\end{align}
\end{subequations}
The expression $g_{i,k}(\mathbf W,\alpha_{i,k},\beta_{i,k})$ is condensed to $g_{i,k}(\mathbf W)$ for conciseness, assuming fixed values for the variables $ \alpha_{i,k} $ and $ \beta_{i,k},k\in\mathcal K,i\in{0,k}$. Problem \eqref{P5} is a convex quadratically constrained quadratic programming (QCQP), which can be solved using standard convex optimization approaches (i.e., interior-point methods) implemented by a certain solver (i.e., SeDuMi) in the CVX toolbox \cite{grant2014cvx}.

\par The approach of using BCD to solve problem (\ref{P3}) and using the standard solvers in CVX to solve problem \eqref{P5} is known as the standard FP method and summarized in Algorithm \ref{alg:AO}. It is also equivalent to the well-known WMMSE approach that has been widely adopted in the state-of-the-art RSMA literature \cite{Mao2022,Hamdi2016,mao2018,Mao2019uni-multicast}. Such equivalence can be easily proved by extending the proof for SDMA in \cite{Shen2018a}, and will also be reflected from the simulation results in Section \ref{simulation}. The standard FP algorithm is ensured to converge to a stationary point $ \mathbf W^\diamond $ of the original problem (\ref{P1}), as indicated by the following Proposition \ref{pro: convergence of AO}.
\begin{proposition}\label{pro: convergence of AO}
Algorithm \ref{alg:AO} is guaranteed to converge to a stationary point of the original problem (\ref{P1}).
\end{proposition}
\par\textit{Proof:} The detailed proof follows the procedure in \cite{Shen2018}.  $ \hfill\blacksquare $ 

\setlength{\textfloatsep}{7pt}	
\begin{algorithm}[t!]
	\caption{Standard FP algorithm to solve problem (\ref{P pureW})}
	\label{alg:AO}
	\textbf{Initilize:} $n \leftarrow 0$, $ \mathrm{WSR}^{[n]}\leftarrow 0 $, $ \epsilon_1 $, $ \mathbf W^{[n]}$\;
	\Repeat{$  |\mathrm{WSR}^{[n]}-\mathrm{WSR}^{[n-1]}|<\epsilon_1 $}{ 
		$ n\leftarrow n+1 $\; 
		Update the auxiliary variables $ \alpha_{i,k}^{[n]} $ and $ \beta_{i,k}^{[n]} $ by (\ref{eq:optalpha}) and (\ref{eq:optbeta}) for $ i=\{0,k\}, k\in\mathcal{K} $\;
		Update $ \mathbf W^{[n]} $ by solving problem (\ref{P5}) with CVX \;}
\end{algorithm}
\subsection{Optimal Beamforming Structure}
\label{subsection OBS}
\par In Subsection \ref{FPM} and Algorithm \ref{alg:AO}, subproblem \eqref{P5} is solved by the standard CVX toolbox. However, such approach is not cost-effective due to the substantial time occupied to parse and canonicalize the original problem instances into standard forms \cite{Shi2015}. In the following, we propose to solve problem (\ref{P5}) based on the KKT optimality. The most compelling contribution of our design is the exploitation of the closed-form solution structure for $\{\mathbf W,y\}$.

\par 
We first analyze problem (\ref{P5}) using the KKT optimality conditions. 
Since (\ref{P5}) is a convex problem and is strictly feasible, problem (\ref{P5}) satisfies the Slater's condition and the strong duality holds \cite{boyd2004convex}. The KKT conditions are therefore sufficient and necessary for the optimal solution. By introducing Lagrangian dual variables $\bm\lambda=[\lambda_1,\cdots,\lambda_K]^T $ and $\mu$ respectively  for (\ref{P5:C1}) and (\ref{P5:C2}), the Lagrange function for problem (\ref{P5}) is defined as 
\begin{equation}
	\label{eq:Lag}
	\begin{aligned}
		\mathcal{L}(\mathbf W,y,&\bm\lambda,\mu)\triangleq\max_{k\in\mathcal K} \{\delta_k\}\cdot y+\sum\limits_{k=1}^K\delta_k g_{k,k}(\mathbf W)\\
		&-\sum\limits_{k=1}^K\lambda_k(y-g_{0,k}(\mathbf W))-\mu(\mathrm{tr}(\mathbf W\mathbf W^H)-P_t).
	\end{aligned}
\end{equation}
The first-order derivatives of (\ref{eq:Lag}) with respect to $\mathbf w_0, \mathbf w_k, k\in\mathcal{K}$, and $y$ are respectively given by
\vspace{-0.5cm}
\begin{equation*}
\begin{aligned}
			\frac{\partial \mathcal{L}}{\partial \mathbf w_0}&=\sum_{j=1}^{K}2\left (d_{0,j} \mathbf h_j -\theta_{0,j}\mathbf h_j  \mathbf h_j^H\mathbf w_0\right )-2\mu\mathbf w_0,\\
			\frac{\partial \mathcal{L}}{\partial \mathbf w_k}&=2d_{k,k}\mathbf h_k-\sum_{j=1}^{K}2\theta_{j,j}\mathbf h_j  \mathbf h_j^H\mathbf w_k-2\mu\mathbf w_k, k\in\mathcal{K},\\
			\frac{\partial \mathcal{L}}{\partial y}&=\max_{k\in\mathcal K} \{\delta_k\}-\sum_{k=1}^K\lambda_k,
\end{aligned}
\end{equation*}
where $ d_{0,j}, \theta_{0,j},\theta_{j,j},\forall j\in\mathcal K$ and $  d_{k,k} ,\forall k\in\mathcal K$
are respectively defined as
\begin{equation*}\label{key}
	\begin{aligned}
		d_{0,j}&\triangleq\sqrt{1+\alpha_{0,j}}\beta_{0,j}\lambda_j,\!\!\!& \theta_{0,j}&\triangleq\lambda_j |\beta_{0,j}|^2,\\
		d_{k,k}&\triangleq\sqrt{1+\alpha_{k,k}}\beta_{k,k}\delta_k,\!\!\!& \theta_{j,j}&\triangleq\delta_j|\beta_{j,j}|^2+\lambda_j |\beta_{0,j}|^2.
	\end{aligned}
\end{equation*}

\par The strong duality of problem \eqref{P5} indicates that there exists a set of optimal Lagrange multiplier $ \lambda_k^\star, k\in\mathcal K $ and $ \mu^\star $, together with optimal primal variables $ \mathbf W^\star,y^\star $, satisfying the KKT conditions of problem \eqref{P5} as follows:
\begin{subequations}
	\label{kkt}   
	\begin{align}
			\label{partial W0} \left(\sum_{j=1}^{K}\theta_{0,j}\mathbf h_j  \mathbf h_j^H\!+\mu\mathbf{I}\right )\mathbf w_0^\star&=\sum_{j=1}^{K}d_{0,j} \mathbf h_j,\\
\label{partial Wk}	\left(\sum_{j=1}^{K}\theta_{j,j}\mathbf h_j  \mathbf h_j^H+\mu\mathbf I\right)\mathbf w_k^\star&=d_{k,k}\mathbf h_k, \forall k\in\mathcal{K},\\
  \label{partial a} \max_{k\in\mathcal K} \{\delta_k\}&=\sum_{k=1}^K\lambda_k^\star\\
		\label{kkt for lambda1}\lambda_k^\star\left(y^\star-g_{0,k}(\mathbf W^\star)\right)&=0,\,\,\,\,\forall k\in\mathcal{K},\\
		\label{kkt for mu}	\mu^\star(\mathrm{tr}(\mathbf W^\star\mathbf W^{\star^H})-P_t)&=0,
	\end{align}
\end{subequations}
where (\ref{partial W0}), (\ref{partial Wk}) and (\ref{partial a}) are the stationary conditions and (\ref{kkt for lambda1}), (\ref{kkt for mu}) refer to the complementary slackness conditions. The primal and dual feasibility conditions are omitted here.

\par  According to \eqref{P5:C1}, it is easy to obtain the optimal solution for $ y $ is given by $ y^\star= \min_{k\in\mathcal K}\{ g_{0,k}\} $. From \eqref{partial W0} and \eqref{partial Wk}, we obtain the closed-form optimal beamforming solution for problem \eqref{P5}, as elucidated in Lemma \ref{Pro: Optimal beamforming of subproblem}.
\begin{lemma}\label{Pro: Optimal beamforming of subproblem}
	The optimal beamforming solution for problem \eqref{P5} is given by
	\begin{subequations}
		\label{up W}
		\begin{align}
			&\mathbf w_0^{\star}\!=\left(\mathbf H\bm\Theta_c(\bm\lambda^\star) \mathbf H^H +\mu^\star \mathbf I\right)^{-1}\mathbf H\mathbf d_c(\bm\lambda^\star),\\
			&\mathbf w_k^{\star} \!=d_{k,k}\left(\mathbf H\bm\Theta_p(\bm\lambda^\star) \mathbf H^H +\mu^\star\mathbf I\right)^{-1} \mathbf h_k, k\in\mathcal K,
		\end{align}
	\end{subequations}
where $ \mathbf H=[\mathbf h_1,\cdots,\mathbf h_K] $ refers to the channel matrix of all users, $\mu^\star$ and $ \bm\lambda^\star $ are the optimal dual variables for problem \eqref{P5}, $\bm\Theta_c(\bm\lambda^\star)$ and $ \bm\Theta_p(\bm\lambda^\star)$ are diagonal matrices with $ \{\theta_{0,1}^\star,\cdots,\theta_{0,K}^\star\} $ and $ \{\theta_{1,1}^\star,\cdots,\theta_{K,K}^\star\} $  respectively along the main diagonals and $ \mathbf d_c(\bm\lambda^\star)=[d_{0,1}^\star,d_{0,2}^\star,\cdots,d_{0,K}^\star]^T $.
\end{lemma}
\subsection{Discussions on the Optimal Beamforming Structure}
The optimal beamforming structure obtained in Subsection \ref{subsection OBS} is in fact the optimal beamforming structure not only for problem \eqref{P5}, but also for problem \eqref{P3} and problem \eqref{P1}. In this subsection, we will obtain and prove the optimal beamforming structure for problem \eqref{P1} using the equivalence relations and specific properties of \eqref{P1} and \eqref{P5}.
\subsubsection{Full Power Consumption Property}
We first introduce Proposition \ref{pro: power constraint} which shows that the transmit power constraints in \eqref{eq:power constraint}, \eqref{P Wc2}, \eqref{P3c3} must hold with equality at any non-zero stationary point. Note that such property has been used in \cite{Park2023} and \cite{Kim2023} without rigorous proof. To the best of our knowledge, this is the first work that prove the full power consumption property for RSMA rigorously.
\begin{proposition}\label{pro: power constraint}
 Any non-zero stationary point of problem \eqref{P1} must satisfy the sum power constraint \eqref{eq:power constraint} with equality.
\end{proposition} 

\textit{Proof:} See Appendix \ref{Appendix: proof of power constraint}.$ \hfill\blacksquare $

\subsubsection{The Optimal Beamforming Structure for Problem \eqref{P1}}

Assuming that we reach an arbitrary non-zero stationary point $ \mathbf W^\diamond $, which also serves as a locally optimal solution for the original problem \eqref{P1}, we have $ \alpha_{i,k}^\diamond=\alpha_{i,k}^\star(\mathbf W^\diamond) $ and $ \beta_{i,k}^\diamond=\beta_{i,k}^\star(\mathbf W^\diamond) $.
Thus, the following Lemma \ref{Lemma:locally optimal} reveals the structure for any locally optimal beamforming solution of problem \eqref{P1}.
\begin{lemma}\label{Lemma:locally optimal}
		Any beamforming solution $ \mathbf W^\diamond $ that is locally optimal for problem \eqref{P1} has the following structure
	\begin{equation}\label{key}
		\begin{split}
			\begin{aligned}
				\mathbf w_c^\diamond&=(\mathbf H\bm\Theta_c^\diamond \mathbf H^H +\mu^\star\mathbf I)^{-1}\mathbf H \mathbf  d_c^\diamond,\\
				\mathbf W_p^\diamond&=(\mathbf H\bm\Theta_p^\diamond \mathbf H^H +\mu^\star\mathbf I)^{-1}\mathbf H\cdot \mathrm{diag}\{\mathbf d_p^\diamond\},
			\end{aligned}
		\end{split}
	\end{equation}
	where $ \bm\lambda^\star,\mu^\star $ are the optimal dual variables and $ \mathbf w_c^\diamond\triangleq\mathbf w_0^\diamond,\mathbf W_p^\diamond\triangleq [\mathbf w_1^\diamond,\cdots,\mathbf w_K^\diamond], \mathbf d_p\triangleq[d_{1,1},\cdots, d_{K,K}]^T,\bm\Theta_c^\diamond= \bm\Theta_c(\bm\lambda^\star,\bm\beta^\diamond),\bm\Theta_p^\diamond= \bm\Theta_p(\bm\lambda^\star,\bm\beta^\diamond),\mathbf d_c^\diamond=\mathbf d_c(\bm\lambda^\star,\bm\alpha^\diamond,\bm\beta^\diamond),\mathbf d_p^\diamond=\mathbf d_p(\bm\alpha^\diamond,\bm\beta^\diamond) $.
\end{lemma}
\textit{Proof:}  In Proposition \ref{pro: convergence of AO}, we already obtain that once we start at a non-zero initial point, the FP-based Algorithm \ref{alg:AO} must converge to a non-zero stationary point $ \mathbf W^\diamond $. Substituting $ \mathbf W^\diamond $ into \eqref{eq:optalpha} and \eqref{eq:optbeta}, the optimal beamforming of problem \eqref{P5} must be the same non-zeros stationary point $ \mathbf W^\diamond $, which implies that all locally optimal beamforming solutions of problem \eqref{P1} share the same beamforming structure with the optimal solution of subproblem \eqref{P5}.  $ \hfill \blacksquare $

\par As the global optimal solution lies within the set of locally optimal solutions \cite{Dong2020}, the FP algorithm would converge to the global optimal solution $ \mathbf W^\circ $ of problem (\ref{P1}) if the initial point $  \mathbf W^{[0]} $ is at the vicinity of $ \mathbf W^\circ $. Therefore, we establish the following
Theorem \ref{Theorem global optimal}.
 \begin{theorem} \label{Theorem global optimal}
 	The optimal beamforming structure for the weighted sum-rate problem in the 1-layer RSMA transmission scheme is given by
\begin{equation}
	\label{global optimal}
	\begin{aligned}
		\mathbf w_c^\circ&=(\mathbf H\bm\Theta_c^\circ \mathbf H^H +\mu^\star\mathbf I)^{-1}\mathbf H \mathbf  d_c^\circ,\\
		\mathbf W_p^\circ&=(\mathbf H\bm\Theta_p^\circ\mathbf H^H +\mu^\star\mathbf I)^{-1}\mathbf H\mathbf D_p^\circ,
	\end{aligned}
\end{equation}  
where $ \bm\lambda^\star,\mu^\star $ are the optimal dual variables and $ \bm\Theta_c^\circ=\bm\Theta_c(\bm\lambda^\star,\bm\beta^\circ), \bm\Theta_p^\circ=\bm\Theta_p(\bm\lambda^\star,\bm\beta^\circ),\mathbf d_c^\circ=\mathbf d_c(\bm\lambda^\star,\bm\alpha^\circ,\bm\beta^\circ),\mathbf d_p^\circ=\mathbf d_p (\bm\alpha^\circ,\bm\beta^\circ), \alpha_{i,k}^\circ=\alpha_{i,k}^\star(\mathbf W^\circ), \beta_{i,k}^\circ=\beta_{i,k}^\star(\mathbf W^\circ) $ and $ \mathbf D_p= \mathrm{diag}\{\mathbf d_p^\circ \} $.
 \end{theorem}

\textit{Proof:} This immediately follows from Lemma \ref{Lemma:locally optimal} by setting the initial point $  \mathbf W^{[0]} $ at the vicinity of $ \mathbf W^\circ $.$ \hfill\blacksquare $ 
\subsubsection{Weighted MMSE Beamforming Structure}
The optimal beamforming structure in Theorem \ref{Theorem global optimal} also indicates that the optimal beamforming vectors for RSMA are weighted MMSE filters, where both the common and private beamformers consist of the following two parts:
\begin{itemize}
	\item The first part is a linear combination of users' channel vectors. To be specific, for the common beamforming vector, the linear combination includes all users' channel vectors, that is, $ \mathbf H\mathbf d_c=\sum_{k=1}^K d_{0,k} \mathbf h_k\triangleq \mathbf h_c$. It is also known as the \textit{group-channel direction}\cite{Dong2020}. The linear combination coefficient $ d_{0,k} $ determines the relative importance of the channel vector $ \mathbf h_k $. For the private beamforming vector, the intended user is only user $ k $, so only $ d_{k,k} $ is non-zero.
	\item The second part is the inverse of the sum of an identity matrix and a weighted channel covariance matrix, where $ \theta_{0,k} $ and $ \theta_{k,k} $ are the weights of each channel $ \mathbf h_k $. This part adjusts the direction of the first part \cite{Bjornson2014}. Generally, $\theta_{0,k}$ is larger than zero, indicating that the design of common beamforming vector $\mathbf w_c$ cannot be recognized as a single group multicast beamforming problem.
\end{itemize}     
This weighted MMSE beamforming structure indicates that the optimal beamforming structure of RSMA combines the optimal unicast and multicast beamforming structures. Specifically, the beamforming of the common stream mirrors that of multicast beamforming, while the beamforming of the private streams mirrors that of unicast beamforming. This is because, from the physical layer transmission perspective, the common stream is multicast to multiple users, while the private streams are unicast to individual users \cite{Mao2022}. However, due to the non-orthogonal nature of RSMA, its optimal beamforming structure more closely resembles to the optimal beamforming structure for non-orthogonal unicast and multicast transmission (NOUM) rather than multi-group multicast beamforming. This
work illustrates how the beamforming structure varies from orthogonal transmission (each user decodes single stream) to non-orthogonal transmission (each user decodes multiple streams), including RSMA, NOUM and NOMA. Moreover, it is important to note that, RSMA fundamentally differs from NOUM since RSMA is originally designed to manage inter-user interference for unicast-only transmission, with the common stream containing parts of the unicast messages for multiple users \cite{Mao2019uni-multicast}.

\subsubsection{Low Dimensional Structure}\label{Sec:low dimension}
Although we have obtained the closed-form expression of the beamforming solution structure, the matrix inverse operation in \eqref{up W} results in a high-dimensional computational complexity of $ \mathcal O(L^3) $. In massive multiple-input multiple-output (MIMO) systems where $L$ is much larger than $K$, computation time increases sharply with $ L $. Fortunately, the optimal beamforming structure presented in Theorem \ref{Theorem global optimal} offers a new perspective on reducing the computational complexity through the method of substitution. However, directly substituting \eqref{global optimal} into the beamforming design problem \eqref{P pureW} leads to an intractable problem with respect to the unknown parameters $\bm\Theta_c, \bm\Theta_p,\mathbf d_c,\mathbf D_c$. 

Instead, we propose to employ an equivalent form of the optimal beamforming structure to avoid inverting the $L$-dimensional matrix, as indicated by the following Proposition \ref{pro:low dimension}.
\begin{proposition}\label{pro:low dimension}
	Any non-zero stationary beamforming solution $ \mathbf W^\diamond $ of problem (\ref{P1}) must has a low dimensional equivalent form, i.e. $ \mathbf W^\diamond=\mathbf H \mathbf Y^\diamond $, with $ \mathbf Y^\diamond\in\mathbb{C}^{K\times (K+1)} $. 
\end{proposition} 
\textit{Proof:} By utilizing the identity $ (\mathbf I+\mathbf A\mathbf B)^{-1}\mathbf A=\mathbf A(\mathbf I+\mathbf B\mathbf A)^{-1} $, any non-zero stationary point stated in the Lemma \ref{Lemma:locally optimal} could be rewritten as
\begin{equation}\label{key}
	\begin{split}
	\mathbf w_c^\diamond&=\mathbf H(\bm\Theta_c^\diamond \mathbf H^H\mathbf H +\mu^\star\mathbf I)^{-1} \mathbf  d_c^\diamond,\\
\mathbf W_p^\diamond&=\mathbf H(\bm\Theta_p^\diamond \mathbf H^H\mathbf H +\mu^\star\mathbf I)^{-1}\cdot \mathrm{diag}\{\mathbf d_p^\diamond\},
	\end{split}
\end{equation}
which implies that there exists a matrix $ \mathbf Y^\diamond\triangleq[(\bm\Theta_c^\diamond \mathbf H^H\mathbf H +\mu^\star\mathbf I)^{-1} \mathbf  d_c^\diamond,(\bm\Theta_p^\diamond \mathbf H^H\mathbf H +\mu^\star\mathbf I)^{-1}\cdot \mathrm{diag}\{\mathbf d_p^\diamond\}]\in\mathbb{C}^{K\times (K+1)} $ such that $ \mathbf W^\diamond=\mathbf H \mathbf Y^\diamond $ always holds.$ \hfill\blacksquare $ 

It is noteworthy that a similar low dimensional property is presented in \cite{rethinking2023}, tailored for SDMA. Proposition \ref{pro:low dimension} has generalized this low dimensional property to the RSMA case.  Our proof is much simpler and more intuitive than the one in \cite{rethinking2023}, as it is built upon a straightforward transformation of the optimal beamforming structure. This offers a novel approach of reducing the computational complexity based on the optimal beamforming structures.
\par By substituting $ \mathbf W=\mathbf H\mathbf Y $ into the WSR maximization problem (\ref{P1}), we obtained the following equivalent low dimensional problem:
\begin{subequations}
	\label{P reduced}
	\begin{align}
		\max\limits_{\mathbf Y,\mathbf c}\,\, &\sum\limits_{k=1}^K \delta_k\left(c_k+\log\left(  1+\frac{|\mathbf f_k^H\mathbf y_k|^2}{\sum_{j=1,j\neq k} |\mathbf f_k^H\mathbf y_j|^2+\sigma_k^2 }\right)\right)\\
		\text{s.t.}\,\,		\label{leq: allocation constraint reduced} &c_k\geq 0,\,\,\, \forall k\in \mathcal{K},\\
		\label{leq: common rate reduced}	&\sum_{i=1}^K c_i\leq \log\left(  1+\frac{|\mathbf f_k^H\mathbf y_0|^2}{\sum_{j=1} |\mathbf f_k^H\mathbf y_j|^2+\sigma_k^2 }\right),\,\,\, \forall k\in\mathcal{K},\\
		\label{eq:power constraint reduced}	& \mathrm{tr}(\mathbf Y\mathbf Y^H\mathbf F)\leq P_{t}.
	\end{align}
\end{subequations}
where $ \mathbf F=[\mathbf f_1,\mathbf f_2,\cdots,\mathbf f_K]=\mathbf H^H\mathbf H\in\mathbf C^{K\times K} $ and $ \mathbf Y=[\mathbf y_0,\mathbf y_1, \cdots,\mathbf y_K] \in\mathbb C^{K\times (K+1)} $. Thus, the dimension of the variables decrease from $ L\times(K+1) $ to $ K\times (K+1) $ and the complexity of matrix inversion decreases from $ \mathcal O(L^3) $ to $ \mathcal O(K^3)  $. Therefore, the derived optimal beamforming structure is advantageous in reducing the computational complexity of RSMA within large-scale massive MIMO networks.
\section{Proposed Optimization Algorithms}\label{Sec:numerical}
 
\par Although we obtain the semi-closed form solution of the optimal $\{\mathbf W^{\star},  y^{\star}\}$ for problem \eqref{P5}, the analytical solutions of the optimal Lagrangian dual variables  $\bm \lambda^{\star}$ and $\mu^{\star}$ embedded in (\ref{up W}) remains challenging to find due to the non-linear dual function obtained by substituting $\{\mathbf W^{\star},  y^{\star}\}$ to (\ref{eq:Lag}).  In this section, we develop a numerical algorithm to calculate $\bm\lambda^\star $ and $ \mu^\star $ without using CVX toolbox. The convergence of the proposed algorithm is then proved rigorously.

\subsection{Hyperplane Fixed Point Iteration}\label{Sec:HFPI}
\par In Section \ref{Sec:optimal}, we already discovered the KKT condition for problem \eqref{P5} in \eqref{kkt}. One classical method to find the Lagrangian dual variables embeded in the optimal beamforming structure is the bisection search, which has been widely adopted in the existing SDMA-assisted networks \cite{Shi2011,Shen2018}.  However, such method cannot be extended for the considered WSR problem of RSMA due to the additional dual variables introduced by the common rate constraints in (\ref{P5:C1}) as well as the resulting KKT conditions \eqref{partial a} and \eqref{kkt for lambda1}, thereby impeding the extension of the existing algorithms for the determination of the optimal values for $ \bm\lambda^\star $ and $ \mu^\star $. Inspired by the classical fixed point iteration (FPI) method \cite{Dong2020,Shi2016,Pham2019,Fan2022} for computing multiple dual variables for problems with multiple nonlinear constraints, we propose an iterative algorithm named hyperplane FPI (HFPI) to find the optimal $ \bm\lambda^\star $ and $ \mu^\star $ in this subsection.

\par Note that the optimal beamforming solution \eqref{up W} of problem \eqref{P5} is completely characterized by the dual variables $ \bm\lambda $ and $ \mu $. Therefore, we denote the beamforming matrix $ \mathbf W $ as a function of $ \bm\lambda $ and $ \mu $, i.e., $\mathbf W=\mathbf W(\bm\lambda,\mu)$, highlighting its dependence on these dual variables. To ease the notation, we further define $ h_{0,k}=g_{0,k}(\mathbf W(\bm\lambda,\mu)) $ and $q= \mathrm{tr}(\mathbf W(\bm\lambda,\mu)\mathbf W^H(\bm\lambda,\mu)) $, respectively. Although we omit the variables, it is important to note that $ h_{0,k} $ and $ q $ are functions with respect to dual variables $ \bm\lambda$ and $\mu $. Suppose that user $ m $ achieves the worst-case common rate in each iteration, i.e., $y=h_{0,m}= \min_{k\in \mathcal{K}} \{h_{0,k}\}$. It is worth noting that the index $ m $ may change in each iteration. We then propose the following HFPI algorithm, which updates the dual variables $\bm\lambda$ and $ \mu $ at iteration $[t+1]$ by
\begin{subequations}
	\label{update_lambda}
	\begin{align}
		\lambda_k^{[t+1]}&=\frac{h_{0,m}^{[t]}+\rho}{h_{0,k}^{[t]}+\rho}\lambda_k^{[t ]}, \,\,\,\, k\neq m, \forall k\in\mathcal{K}\\
		\lambda_m^{[t+1]}&=\lambda_m^{[t]}\!+\!\sum_{k=1}^K\!\left(\!1\!-\!\frac{h_{0,m}^{[t]}+\rho}{h_{0,k}^{[t]}+\rho}\right)\lambda_k^{[t]},\\
		\mu^{[t+1]}&=\frac{q^{[t]}+\rho}{P_t+\rho}\mu^{[t]},
	\end{align}
\end{subequations}
where constant $ \rho\geq 0 $ is employed to enhance convergence stability by effectively reducing the step size. Thus, the dual variable $ \bm\lambda $ always meets constraint (\ref{partial a}) in each iteration. The detail procedure for deriving \eqref{update_lambda} is provided in the Appendix \ref{Appendix: proof of FP}.
%

\subsection{Overall Optimization Framework for Problem \eqref{P1}}

Combining the HFPI algorithm proposed in Section \ref{Sec:HFPI} and the FP framework introduced in Section \ref{FPM}, we obtain the overall optimization algorithm, named FP-HFPI, to solve problem \eqref{P1}. The details of FP-HFPI are illustrated in Algorithm \ref{alg:proposed}. Starting with a non-zero feasible beamforming matrix $ \mathbf W^{[0]} $, we first calculate the auxiliary variables $ \alpha_{i,k}^{[n]} $ and $ \beta_{i,k}^{[n]} $ at Step 4 in iteration $ [n] $. With the fixed auxiliary variables, the bemaforming matrix $ \mathbf W^{[t]} $ and dual variables $ \bm\lambda^{[t]},\mu^{[t]} $ are updated iteratively according to \eqref{up W} amd \eqref{update_lambda} at Step 8 and 9, respectively. The fixed points $ \bm\lambda^{[n]},\mu^{[n]} $ are obtained until the convergence of the inner-layer HFPI algorithm. Then, the beamforming matrix $ \mathbf W^{[{n}]} $ is updated based on the obtained $ \bm\lambda^{[n]},\mu^{[n]} $, which are also the optimal dual variables for problem \eqref{P5}, as proven in the following subsection. The outer-layer FP framework ensures that Algorithm \ref{alg:proposed} converges to a stationary point of the original problem \eqref{P1} as stated in Proposition \ref{pro: convergence of AO}.

\setlength{\textfloatsep}{7pt}	
\begin{algorithm}[t!]
	\caption{Proposed FP-HFPI algorithm to solve problem (\ref{P1})}
	\label{alg:proposed}
	\textbf{Initilize:} $n \leftarrow 0$, $ \mathrm{WSR}^{[n]}\leftarrow 0 $, $ \epsilon_1 $, $ \epsilon_2 $, $ \bm\lambda^{[n]} $, $ \mu^{[n]}$, $ \mathbf W^{[n]}$\;
	\Repeat{$ |\mathrm{WSR}^{[n]}-\mathrm{WSR}^{[n-1]}|<\epsilon_1 $}{ 
		$ n\leftarrow n+1 $\;
		Update the auxiliary variables $ \alpha_{i,k}^{[n]} $ and $ \beta_{i,k}^{[n]} $ by (\ref{eq:optalpha}) and (\ref{eq:optbeta}) for $ i=\{0,k\}, k\in\mathcal{K} $\;
		$t \leftarrow 0$\;
		\Repeat{$ \|\bm\lambda^{[t]}-\bm\lambda^{[t-1]}\|+|\mu^{[t]}-\mu^{[t-1]}|<\epsilon_2  $}{$ t\leftarrow t+1 $\;
        Update $ \mathbf W^{[t]} $ by (\ref{up W})\;
        Update $ \bm\lambda^{[t]} $ and $ \mu^{[t]}  $  by (\ref{update_lambda})\;}
    Update $ \bm\lambda^{[n]}=\bm\lambda^{[t]} $ and $ \mu^{[n]}=\mu^{[t]} $\;
    Update $ \mathbf W^{[n]} $ with $ \bm\lambda^{[n]}$ and $\mu^{[n]}$ by (\ref{up W})\;
}
\end{algorithm}
\subsection{Convergence Analysis}
\par The proposed FP-HFPI algorithm involves two iterations. The convergence of outer layer iteration mirrors the standard FP algorithm in Algorithm \ref{alg:AO}, which has been proved in Proposition \ref{pro: convergence of AO}. In this subsection, we delve into the convergence analysis of the inner layer iteration, which is the HFPI algorithm for computing the optimal dual variables.

To prove the convergence of $ \bm\lambda,\mu $, we undertake the following two steps:
 \begin{itemize}
 	\item \textit{Step 1 - Uniqueness Analysis:} We first prove that there exists and only exists one set of $ \bm\lambda^\star $ and $ \mu^\star $ satisfying equation \eqref{partial a}, \eqref{kkt for lambda1} and \eqref{kkt for mu}.
 	 	\item \textit{Step 2 - Convergence Analysis:} Next, we prove that the L1 norm of the discrepancy between $ \bm\lambda^\star,\mu^\star $ and the solution obtained by HFPI converges to zero, implying that HFPI converges to $ \bm\lambda^\star,\mu^\star $.
 \end{itemize}
\subsubsection{Uniqueness Analysis}
Although it is challenging to solve the non-linear equations in \eqref{kkt} directly, we remain able to obtain some useful information about the solution structure from the KKT conditions. Consider that there exists an optimal dual vector $ \bm\lambda^\star=[\lambda_1^\star,\cdots,\lambda_K^\star]^T $ satisfying the KKT condition \eqref{kkt for lambda1} as follows
\begin{equation}\label{KKT for opt lambda}
\begin{aligned}
		\lambda_k^\star (y-h_{0,k}^\star)&=0,\,\,\,\forall k\in\mathcal K.
\end{aligned}
\end{equation} 
All $ \lambda_k^\star,\forall k\in\mathcal K $ can be divided into two subsets, one subset is $ \bm\lambda_N=\{\lambda_k^\star|\lambda_k^{\star}\neq 0,\forall k\in\mathcal K\}$ and the collection of their indices is denoted as $ \mathcal{N} $. The other subset is $ \bm\lambda_Z=\{\lambda_k^\star|\lambda_k^{\star}=0,\forall k\in\mathcal K\} $ and the collection of their indices is denoted as $  \mathcal{Z} $. Obviously, we have  $ \mathcal{N}\cup\mathcal{Z}=\mathcal{K},\mathcal{N}\cap\mathcal{Z}=\emptyset $. It should be clear that $ \bm\lambda_N $ and $ \bm\lambda_Z $ are defined to help us understand the following corollary \ref{corollary equality}.
 \begin{corollary} \label{corollary equality}
	All users in $ \mathcal N $ have the same achievable rate to decode stream $ s_0 $, which also is equal to the common rate, i.e., 
	\begin{equation}\label{non zero lambda}
		y^\star=h_{0,n}^\star,\forall n\in\mathcal N.
	\end{equation} 
 The remaining users in $ \mathcal Z $ always have an achievable rate greater than the common rate, i.e., 
\begin{equation}
\label{zero lambda}	y< h_{0,z}^\star,\forall z\in\mathcal{Z}.
\end{equation}
According to Proposition \ref{pro: power constraint} which indicates that $\mu^\star>0  $, the complementary slackness condition \eqref{kkt for mu} reduces to \begin{equation}\label{key}
	q^\star=P_t.
\end{equation}
\end{corollary}
%


With the help of above Corollary \ref{corollary equality}, we obtain the following Proposition \ref{pro: only}.
 \begin{proposition} \label{pro: only}
 There exists and only exists one unique fixed point $ \bm\lambda^\star,\mu^\star $ satisfying \eqref{partial a}, \eqref{kkt for lambda1} and \eqref{kkt for mu}.
 \end{proposition}
\par\textit{Proof:} See Appendix \ref{Appendix:proof of unieq}.

\subsubsection{Convergence Analysis}

\par  For the dual problem of \eqref{P5} which minimizes the Lagrangian function defined in \eqref{eq:Lag}, the dual variable $ \lambda_k $ and $ \mu $ could be interpreted as the optimization weight associated with the rate $ h_{0,k} $ and transmit power $ q $. Then, we are able to obtain the following Lemma \ref{lemma: monotone}.
\begin{lemma}\label{lemma: monotone}
	$ h_{0,k} $ is a monotonically increasing function with respect to $ \lambda_k $ and $q $ is monotonically decreasing with respect to $ \mu $.
\end{lemma}

\textit{Proof:} See Appendix \ref{Appendix:monotic}.$ \hfill \blacksquare $ 

Lemma \ref{lemma: monotone} provides the insight of relative location between $ \lambda_k^{[t]},\mu^{[t]} $ and $ \lambda_k^\star,\mu^\star $.
Then the convergence of the proposed HFPI is guaranteed by the following Lemma \ref{convergence of dual}.
\begin{lemma}\label{convergence of dual}
	Assume that $ \bm\lambda^\star,\mu^\star $ are the optimal dual variables for problem \eqref{P5}. Then there exists a finite constant $ \rho<\infty $, such that 
\begin{equation}
	\lim_{t\rightarrow \infty}\|\bm\lambda^{[t]}-\bm\lambda^{\star}\|_1=0,\lim_{t\rightarrow \infty}\|\mu^{[t]}-\mu^{\star}\|_1=0.
\end{equation}
\end{lemma} 

\textit{Proof:} According to the relative value between $ \lambda_k^{[t]} $ and $ \lambda_k^\star $, we divide $ \bm\lambda^{[t]} $ into the following three parts: $ \bm\lambda_L^{[t]}=\{\lambda_k^{[t]}|\lambda_k^{[t]}\leq\lambda_k^\star,\forall k\in\mathcal K, k\neq m\} $, $ \bm\lambda_R^{[t]}=\{\lambda_k^{[t]}|\lambda_k^{[t]}>\lambda_k^\star,\forall k\in\mathcal K\} $, and the rest is $ \lambda_m^{[t]} $. Specially, $ \lambda_m^{[t]}<\lambda_m^{\star} $ since $ h_{0,m}=y<y^\star\leq h_{0,m}^\star $. Further, denote the collections of indices for $ \bm\lambda_L^{[t]} $ and $ \bm\lambda_R^{[t]}$ as $ \mathcal{L} $ and $ \mathcal{R} $, respectively. \begin{figure}[htbp]
	\centering
	\includegraphics[width=0.44\textwidth]{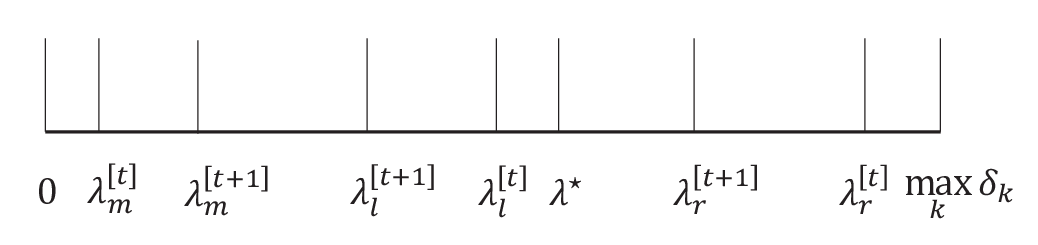}
	\caption{The relative position of $ \lambda_k $ to its fixed point $ \lambda_k^\star $.}
	\label{fig1}
\end{figure} Fig. \ref{fig1} shows their relative location between $ [0,1] $. In each iteration, we assume that their relative positions to their fixed point remains unchanged, i.e., there exists a finite constant $ \rho $ satisfying
\begin{subequations}	\label{ine}
	\begin{align}		
	&\lambda_r^\star\leq\lambda_r^{[t+1]},\forall r\in\mathcal{R}, \\
		\label{ine 2}	&\lambda_m^{[t+1]}\leq\lambda_m^\star,\\
		&\mu^\star\leq \mu^{[t+1]}.
	\end{align}
\end{subequations}
The step size between two consecutive iterations is given by
\begin{equation*}
	\begin{aligned}
		&\|\bm\lambda^{[t+1]}-\bm\lambda^\star\|_1-\|\bm\lambda^{[t]}-\bm\lambda^\star\|_1+|\mu^{[t+1]}-\mu^\star|_1-|\mu^{[t]}-\mu^\star|_1\\
		&=\sum_{l\in\mathcal{L}}\left(\lambda_l^{[t]}-\lambda_l^{[t+1]}\right)+\sum_{r\in\mathcal{R}}\left(\lambda_r^{[t+1]}-\lambda_r^{[t]}\right)\\
		&+\left(\lambda_m^{[t]}-\lambda_m^{[t+1]}\right)+(\mu^{[t+1]}-\mu^{[t]})\\ &=\sum_{l\in\mathcal{L}}\left(\frac{h_{0,l}^{[t]}-h_{0,m}^{[t]}}{h_{0,l}^{[t]}+\rho}\right)\lambda_l^{[t]} -\sum_{r\in\mathcal{R}} \left(\frac{h_{0,r}^{[t]}- h_{0,m}^{[t]}}{h_{0,r}^{[t]}+\rho}\right)\lambda_r^{[t]}\\ &-\sum_{k\in\mathcal{K}}\left(\frac{h_{0,k}^{[t]}- h_{0,m}^{[t]}}{h_{0,k}^{[t]}+\rho}\right)\lambda_k^{[t]}+\frac{q^{[t]}-P_t}{P_t+\rho}\mu^{[t]}\\
		&=-2\sum_{r\in\mathcal{R}}\left(\frac{h_{0,r}^{[t]}-h_{0,m}^{[t]}}{h_{0,r}^{[t]}+\rho}\right)\lambda_r^{[t]}+\frac{q^{[t]}-P_t}{P_t+\rho}\mu^{[t]}\\
		&\leq 0
	\end{aligned}
\end{equation*}
The first equality follows \eqref{update_lambda} and \eqref{ine}, and last inequality achieves an equality sign when $ \bm\lambda^{[t]}=\bm\lambda^\star, \mu^{[t]}=\mu^\star $, which ensures that the algorithm converges to the fixed point $ \bm\lambda^\star,\mu^\star$.

Let user $ d $ be the user that achieves the maximum common rate in each iteration, i.e., $ h_{0,d}=\max_{k\in\mathcal{K}}\{h_{0,k}\} $, which contrasts with user $ m $ that achieves the minimum achievable common rate. {Based on \eqref{update_lambda}, the inequalities in (\ref{ine}) are expressed as
\begin{subequations}
	\begin{align}
	\label{rhor}	\lambda_r^\star&\leq\frac{h_{0,m}^{[t]}+\rho}{h_{0,r}^{[t]}+\rho},\forall r\in\mathcal{R},\\
		\label{rhom}\lambda^\star_m&\geq \lambda_m^{[t]}\!+\!\sum_{k=1}^K\!\left(\!1\!-\!\frac{h_{0,m}^{[t]}+\rho}{h_{0,d}^{[t]}+\rho}\right)\lambda_k^{[t]},\\
		&\geq \label{mus}\lambda_m^{[t]}\!+\!\sum_{k=1}^K\!\left(\!1\!-\!\frac{h_{0,m}^{[t]}+\rho}{h_{0,k}^{[t]}+\rho}\right)\lambda_k^{[t]},\\
		\mu^\star&\leq \frac{q^{[t]}+\rho}{P_t+\rho}\mu^{[t]}.
	\end{align}
\end{subequations}
We further simplify \eqref{rhom} to  
\begin{equation}\label{rhom2}
	\lambda_m^\star\geq \lambda_m^{[t]}+\left(1-\frac{h_{0,m}^{[t]}+\rho}{h_{0,d}^{[t]}+\rho}\right)\cdot \max_{k\in\mathcal K}\{\delta_k \}.
\end{equation}
Therefore, \eqref{rhor}, \eqref{mus} and \eqref{rhom2} are able to be rewritten as
\begin{subequations}\label{optimal rho}
	 	\begin{align}
	 		&\rho\geq\frac{\lambda_r^\star h_{0,r}^{[t]}-\lambda_r^{[t]} h_{0,m}^{[t]}}{\lambda_r^{[t]}-\lambda_r^\star},\forall r\in\mathcal{R}, \\
	&\rho\geq\frac{h_{0,d}^{[t]}-h_{0,m}^{[t]}}{\lambda_m^\star-\lambda_m^{[t]}}\max_{k\in\mathcal K}\{\delta_k \}-h_{0,d}^{[t]},\\
	&\rho \geq \frac{\mu^\star P_t-\mu^{[t]}q^{[t]}}{\mu^{[t]}-\mu^\star}.
	 	\end{align}
\end{subequations}
The right-hand sides of all inequalities in \eqref{optimal rho} are continuously differentiable and their limits exist when $ \bm\lambda,\mu $ converges to $ \bm\lambda^\star,\mu^\star $. Thus, the convergence of $ \bm\lambda,\mu $ is established.} $ \hfill \blacksquare $ 

\par We also show that the proposed HFPI algorithm not only ensures convergence but also converges specifically to the optimal beamforming solution of the convex subproblem (\ref{P5}) in the following Proposition \ref{pro:convergence of HFPI}.
\begin{proposition}\label{pro:convergence of HFPI}
	{The proposed HFPI algorithm is guaranteed to converge to the optimal beamforming solution of problem \eqref{P5}. }
\end{proposition}

\par \textit{Proof:} {Based on Proposition \ref{pro: only} and Lemma \ref{convergence of dual}, the dual variables are guaranteed to converge to the fixed point of the KKT conditions \eqref{kkt}. As problem (\ref{P5}) is convex and strong duality holds, the fixed point that satisfies the KKT conditions is the optimal dual solution. By substituting the optimal dual variables into (\ref{up W}), we obtain the optimal solution of the convex subproblem (\ref{P5}), thereby completing the proof.}  $ \hfill \blacksquare $ 

\subsection{Computational Complexity Analysis} 

For the proposed Algorithm \ref{alg:proposed}, the computational complexity for per iteration is dominated by updating the beamforming matrix (i.e., line 8 of Algorithm \ref{alg:proposed}) with the order of $ \mathcal{O}(KL^3) $. Hence, the total complexity order of Algorithm \ref{alg:proposed} is $\mathcal{O}(I_{\bm\lambda}I_{\mathbf W}KL^3)  $, where $ I_{\bm\lambda}$ and $I_{\mathbf W} $ represent the number of iterations for inner loop to find the optimal dual variables and outer loop to reach a stationary point of problem \eqref{P1}, respectively. 


\section{Extension to Imperfect CSIT Scenarios}\label{Sec:imperfect}
{ In this section, we extend our proposed optimization framework to solve the WSR problem of RSMA under imperfect CSIT scenario. The joint optimization of beamforming and common rate allocation for the stochastic WSR maximization problem of RSMA in imperfect CSIT has been widely studied \cite{Hamdi2016,Li2020,Fu2020,Dai2016,Park2023}. Generally, there are two primary approaches used in existing works for addressing the stochastic nature of such problem: the first one approximates the distribution of the channel estimation error by a large number of channel samples, also known as sample average approximation (SAA) \cite{Hamdi2016}, the other one establishes the lower bound of the stochastic rate expressions \cite{Park2023}. In the following, we take the lower-bound approximation as an example and show the extension of our proposed optimization framework for addressing the WSR maximization problem of RSMA under imperfect CSIT scenarios. 
\subsection{Imperfect CSIT Model}

\par Assuming that the channel vector $ \mathbf h_k $ follows spatially correlated Gaussian distribution, i.e., $ \mathbf h_k\sim \mathcal{CN}(\mathbf 0,\mathbf R_k)$, where $ \mathbf R_k=\mathbb{E}[\mathbf h_k\mathbf h_k^H] $ is the channel covariance matrix constructed by the one-ring model \cite{onering2013}.  By employing the Karhunen-Loeve factorization \cite{onering2013}, we further express $ \mathbf h_k $ as $ \mathbf h_k=\mathbf U_k\bm\Lambda^{\frac{1}{2}}_k\mathbf g_k $, where $ \bm\Lambda_k $ is a diagonal matrix containing the eigenvalues of $ \mathbf R_k $, $ \mathbf U_k $ contains the eigenvectors of $ \mathbf R_k $ corresponding to the eigenvalues in $ \bm\Lambda_k $, and $ \mathbf g_k $ is an independent and identically distributed (i.i.d) vector with entries drawn from $ \mathcal{CN}(0,1) $. Considering a frequency division duplex (FDD) MIMO scenario with limited feedback, the estimated CSIT $\hat{\mathbf h}_k  $ is modeled as
\begin{subequations}
	\begin{align}
		\hat{\mathbf h}_k&=\mathbf h_k-\mathbf e_k\\
		&=\mathbf U_k\bm\Lambda^\frac{1}{2}(\sqrt{1-\kappa^2}\mathbf g_k+\kappa\mathbf g_{e,k}),
	\end{align}
\end{subequations} 
where $ \mathbf e_k $ refers to the quantization error vector and $ \mathbf g_{e,k} $ is an i.i.d vector with entries drawn from $ \mathcal{CN}(0,1) $. The quality of CSIT is quantified by the parameter $ \kappa \in [0,1]$. When $ \kappa=0 $, the channel estimate at the transmitter is perfect, i.e., $ \hat{\mathbf h}_k=\mathbf h_k $. With the above assumption, the covariance matrix for the quantization error $ \mathbf e_k $ is given by
 \begin{equation}
 \label{error co}	\bm\Phi_k=\mathbb{E}[\mathbf e_k\mathbf e_k^H]=\mathbf U_k\bm\Lambda^\frac{1}{2}(2-2\sqrt{1-\kappa^2})\bm\Lambda^\frac{1}{2}\mathbf U_k^H .
 \end{equation}
  
\subsection{Rate Lower Bound and Problem Formulation}

\par Based on the imperfect CSIT model as described above, the instantaneous rate $r_{i,k}$ corresponding to each channel estimate may not decodable. To ensure achievability, we follow \cite{Park2023,Kim2023} by deriving a lower bound for each rate expression $ r_{i,k} $. To be specific, by treating the CSIT error as an independent Gaussian noise, a lower bound $ \widetilde{r}_{i,k} $ for the rate $ r_{i,k} $ is obtained by 
\begin{equation}\label{key}
r_{i,k}\geq\widetilde{r}_{i,k}=\log\left(1+\tilde{\gamma}_{i,k}\right),
\end{equation}
where 	$ \tilde{\gamma}_{i,k}=  |\hat{\mathbf h}_k^H\mathbf w_i|^2/ (\sum_{j=1,j\neq i}^K |\hat{\mathbf h}_k^H\mathbf w_j|^2+\sum_{j\in\mathcal{K}\cup\{i\}}\mathbf w_j^H\bm\Phi_k\mathbf w_j+\sigma_k^2). $ More details about the derivation of the lower bound can be found in \cite{Park2023,Choi2020,Kim2023}. It should be noted that although the lower bound $ \tilde{\gamma}_{i,k} $ introduces additional quadratic terms in the denominator, they do not bring additional optimization challenges compared to the original achievable SINR $ \gamma_{i,k} $.

\par Based on the above lower bound of the instantaneous rate, we formulate the WSR maximization problem for RSMA based on the instantaneous estimated CSIT and the CSIT error covariance, which is given as follows:
\begin{subequations}
	\label{P2}
	\begin{align}
		\max\limits_{\mathbf W,\mathbf c}\,\, &\sum\limits_{k=1}^K \delta_k(c_k+\tilde{r}_{k,k})\\
		\text{s.t.}\,\,		 &c_k\geq 0,\,\,\, \forall k\in \mathcal{K},\\
		&\sum_{i=1}^K c_i\leq \tilde{r}_{0,k},\,\,\, \forall k\in\mathcal{K},\\
		& \mathrm{tr}(\mathbf {WW}^H)\leq P_{t}.
	\end{align}
\end{subequations}
 Problem \eqref{P2} is more general formulation than problem \eqref{P1}. When $ \kappa $ is set to zero, the channel CSIT error covariance matrices becomes zeros matrices. Consequently, problem \eqref{P1} reduces to problem \eqref{P1}.
\begin{remark}\label{remark 1}
	 In this work, our focus is on solving problem \eqref{P1}. On one hand, once problem \eqref{P1} is efficiently solved, we can easily extend our optimization framework to solve problem \eqref{P2} under imperfect CSIT scenarios, as the fundamental challenges between the two problems remain the same after transforming the stochastic imperfect CSIT optimization problem into a deterministic one. On the other hand, the lower-bound approximation approach described in this section is one representative approach to handle the stochastic nature of the problem. Other approaches, such as SAA \cite{Hamdi2016}, exist as well.  Our proposed optimization framework for problem (2) with perfect CSIT can also be applied to tackle problems utilizing these alternative imperfect CSIT methods. 
\end{remark} 

\subsection{Optimization Framework to Problem \eqref{P2}}

Problem (\ref{P2}) can be efficiently solved with some minor modifications to the proposed optimization framework. Specifically, by applying the FP method to problem (\ref{P2}), the optimal auxiliary variables $ \tilde{\alpha}_{i,k}^\star $ and $ 	\tilde{\beta}_{i,k}^\star $ are given by
\begin{equation}\label{imFP}
	\begin{split}	
		\tilde{\alpha}_{i,k}^\star&=\frac{ |\hat{\mathbf h}_k^H\mathbf w_i|^2 }{ \sum_{j=1,j\neq i}^K |\hat{\mathbf h}_k^H\mathbf w_j|^2+\sum_{j\in\mathcal{K}\cup\{i\}}\mathbf w_j^H\bm\Phi_k\mathbf w_j+\sigma_k^2},\\
		\tilde{\beta}_{i,k}^\star&=\frac{\sqrt{1+\tilde{\alpha}_{i,k}}\hat{\mathbf{h}}_{k}^{H}\mathbf w_i}{\sum_{j\in\mathcal{K}\cup\{i\}}|\hat{\mathbf h}_k^{H}\mathbf w_j|^2+\sum_{j\in\mathcal{K}\cup\{i\}}\mathbf w_j^H\bm\Phi_k\mathbf w_j+\sigma_k^2}.
	\end{split}
\end{equation}
Then, we are able to focus on the subproblem with respect to $ \{\mathbf W,y\} $. Based on the corresponding KKT conditions, the optimal beamforming vectors for the common and private streams are given by
\begin{equation}\label{imKKT}
	\begin{split}
			& \tilde{\mathbf w}_0^{\star}=\left(\sum_{j=1}^{K}\tilde{\theta}_{0,j}(\hat{\mathbf h}_j  \hat{\mathbf h}_j^H+\bm\Phi_k)+\mu\mathbf I\right)^{-1}\sum_{j=1}^K \tilde{d}_{0,j} \hat{\mathbf h}_j,\\
 	&\tilde{\mathbf w}_k^{\star} =\left(\sum_{j=1}^K \tilde{\theta}_{j,j}(\hat{\mathbf h}_j \hat{\mathbf h}_j^H+\bm\Phi_k) +\mu\mathbf I  \right)^{-1}\tilde{d}_{k,k}\hat{\mathbf h}_k,\quad\forall k\in\mathcal{K}.
	\end{split}
\end{equation}
where $ \tilde{d}_{0,j}, \tilde{\theta}_{0,j},\tilde{\theta}_{j,j},\forall j\in\mathcal K$ and $  \tilde{d}_{k,k} ,\forall k\in\mathcal K$
are respectively defined as
\begin{equation*}\label{key}
	\begin{aligned}
		\tilde{d}_{0,j}&\triangleq\sqrt{1+\tilde{\alpha}_{0,j}}\tilde{\beta}_{0,j}\lambda_j,& \tilde{\theta}_{0,j}&\triangleq\lambda_j |\tilde{\beta}_{0,j}|^2,\\
		\tilde{d}_{k,k}&\triangleq\sqrt{1+\!\tilde{\alpha}_{k,k}}\tilde{\beta}_{k,k}\delta_j,& \tilde{\theta}_{j,j}&\triangleq \delta_{j,j}|\tilde{\beta}_j|^2+\lambda_j |\tilde{\beta}_{0,j}|^2.
	\end{aligned}
\end{equation*} 
The update of dual variables $ \lambda^\star,\mu^\star $ directly follows the HFPI algorithm proposed in Section \ref{Sec:HFPI} . The entire procedure for solving problem (\ref{P2}) follows the same steps outlined in Algorithm \ref{alg:proposed} by simply changing the update of the auxiliary variables based on (\ref{imFP}) and the update of the beamforming vectors based on (\ref{imKKT}). We omit the details to eliminate redundancy.}

\section{Simulation Results}\label{simulation}\label{Sec:simulation}
In this section, we evaluate the performance of the proposed  FP-HFPI algorithm. Two classical methods are compared as baselines , namely WMMSE and SCA. It should be noted that both of them require a standard optimization solver in toolbox such as CVX \cite{grant2014cvx} and Yalmip \cite{Lofberg2004} to solve a convex subproblem in each iteration. Another CVX-free approach proposed in \cite{Park2023}, named GPI is also considered as a baseline scheme. All schemes considered in the simulation are summarized as follows:
\begin{itemize}
	\item \textbf{WMMSE}: The classical WMMSE approach transforms the non-convex WSR problem into a block-wise convex problem by the rate-WMMSE relation and solves each block iteratively until convergence \cite{Hamdi2016}. 
	\item \textbf{SCA}: The SCA method solves the non-convex WSR problem via constructing a sequence of convex surrogates, which are solved iteratively until convergence \cite{Mao2019uni-multicast}.
	\item \textbf{FP}: The standard FP method is specified in Algorithm \ref{alg:AO}. It transforms the WSR problem into an equivalent block-wise convex problem by the Lagrangian dual and quadratic transform and solves each block iteratively until convergence \cite{Shen2018}.
	\item \textbf{GPI}: The GPI method uses the LogSumExp function to approximate the common rate into a smooth form and solves the approximated problem via its first-order optimality condition by GPI \cite{Park2023}.
	\item \textbf{FP-HFPI}: The proposed FP-HFPI method  specified in Algorithm \ref{alg:proposed}.
	\item \textbf{FP-HFPI-s}: This corresponds to a special instance of the proposed FP-HFPI algorithm by setting the tolerance of HFPI to a value larger than $ \max_{k\in\mathcal K}\{\delta_k \} $. It is a heuristic but effective approach we discovered from extensive simulation.
	\item \textbf{RFP-HFPI}: This corresponds to a special instance of the proposed FP-HFPI algorithm specified in Section \ref{Sec:low dimension}. It transforms the original WSR problem into an equivalent low dimensional form \eqref{P reduced}, which is then solved by FP-HFPI.
	\item \textbf{RFP-HFPI-s}: This corresponds to a special instance of the proposed RFP-HFPI algorithm by setting the tolerance of HFPI to a value larger than $ \max_{k\in\mathcal K}\{\delta_k \} $.
\end{itemize}
The computational complexity of the above algorithms for solving the WSR problem \eqref{P1} are illustrated in Table \ref{tab:complexity} when $ K $ and $ L $ increases with the same order. It should be clear that WMMSE, SCA, and FP requires standard solvers to solve a convex subproblem in each iteration, while the rest four algorithms are toolbox free.
\begin{table}[t!]
	\begin{center}
		\caption{Computational Complexity Comparison for Different Algorithms to Solve Problem \eqref{P1}.}
		\label{tab:complexity}
		\begin{tabular}{ccc}
			\toprule
			\textbf{Algorithm}& 	\textbf{Reference} & \textbf{Computational Complexity}  \\
			\hline\hline
			\textbf{WMMSE}& \cite{Hamdi2016}&$ \mathcal O(I_{\mathbf W}[KL]^{3.5})$  \\
			\hline
			\textbf{SCA}& \cite{Mao2019uni-multicast}&$\mathcal O(I_{\mathbf W}[KL]^{3.5}) $  \\
			\hline
			\textbf{Standard FP}& \cite{Shen2018}&$\mathcal O(I_{\mathbf W}[KL]^{3.5}) $\\
			\hline
			\textbf{GPI}& \cite{Park2023}&$\mathcal O(I_{\mathbf W}KL^3) $\\
			\hline
			\textbf{FP-HFPI}&Proposed & $\mathcal{O}(I_{\bm\lambda}I_{\mathbf W}KL^3)  $\\
			\hline
			\textbf{FP-HFPI-s}&Proposed& $\mathcal{O}(I_{\mathbf W}KL^3)  $\\
			\hline
			\textbf{RFP-HFPI}&Proposed & $\mathcal{O}(K^2L+I_{\bm\lambda}I_{\mathbf W}K^4)  $\\
			\hline
			\textbf{RFP-HFPI-s}&Proposed& $\mathcal{O}(K^2L+I_{\mathbf W}K^4)   $\\
			\bottomrule
		\end{tabular}
	\end{center}
\end{table}

\par We first consider a classic MISO BC with $ L=4 $ and $ K=4 $. The channel of user $ k $ is generated i.i.d. as $ \mathbf h_k\sim\mathcal{CN}(\mathbf 0,\mathbf I) $ and the noise variance at user $ k $ is set to $ \sigma_k^2=1 $ so that the transmit signal-to-noise (SNR) SNR $\triangleq P_t/\sigma_k^2  $ is numerically equal to the transmit power. For Algorithm \ref{alg:proposed}, we set $ \rho=0.5 $ and the stopping tolerance $ \epsilon_2=10^{-3}  $ to control the convergence accuracy of the inner HFPI. We also set $ \epsilon_2=10^0 $ as a special case, namely FP-HFPI-s, since it could maintain almost the same performance as FP-HFPI while further reducing the computation time. The convergence tolerance for the outer FP approach is $ \epsilon_1=10^{-4} $. Without loss of generality, we set $ \delta_1=\delta_2=\cdots=\delta_K=1 $ in most of our simulations. All simulation results are averaged over $ 100 $ random channel realizations.
\begin{figure*}
	\begin{center}
		\begin{minipage}{0.3\textwidth}
		\includegraphics[width=1\linewidth]{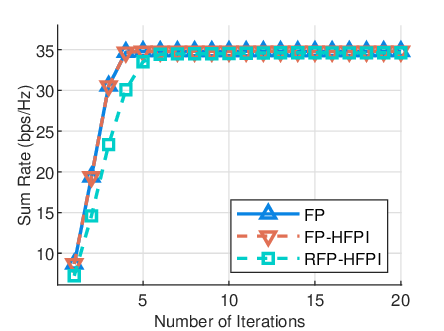}
		\caption{Convergence of the proposed algorithms.}
		\label{fig2}
		\end{minipage}
		\begin{minipage}{0.3\textwidth}
			\includegraphics[width=1\linewidth]{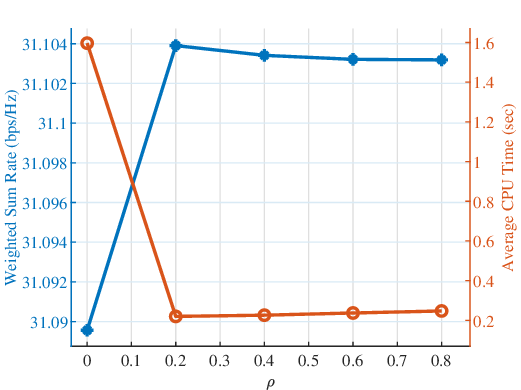}
			\caption{Performance with the increase of $ \rho $. }
			\label{fig:rho}
		\end{minipage}
		\begin{minipage}{0.3\textwidth}
		\includegraphics[width=1\linewidth]{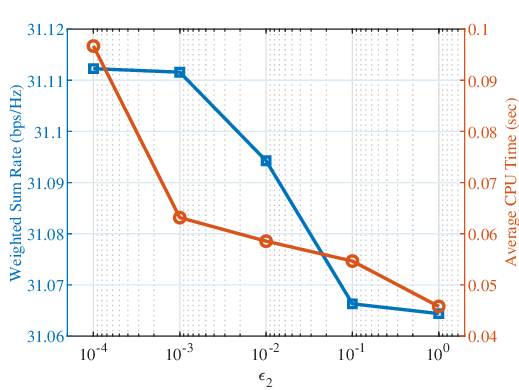}
		\caption{Performance with the increase of $ \epsilon_2$. }
		\label{fig:epsilon}
		\end{minipage}
	\end{center}
\end{figure*}

%
%
%
%
%
\subsection{Convergence Behavior and Parameter Effect}
\par  We first check the convergence behavior of the standard FP algorithm, the proposed FP-HFPI algorithm and it's special case, the RFP-HFPI algorithm. Fig.$\, \ref{fig2} $ showcases that the proposed FP-HFPI and RFP-HFPI achieve nearly identical WSR when compared with the standard FP algorithm, which aligns with the theoretical analysis in Section \ref{Sec:optimal} and \ref{Sec:numerical}.

\par  
Fig.$\, \ref{fig:rho} $ demonstrates the effects of parameter $ \rho $ on the WSR and average CPU time performance of the FP-HFPI when the transmit SNR is $ 30$ dB. To ensure a stable convergence, $ \rho $ should be chosen to satisfy \eqref{optimal rho}. Identifying the smallest $ \rho $ analytically poses a  challenge, but finding an appropriate $ \rho $ numerically is a straightforward task. It is observed that as the constant $ \rho $ increases, the performance of FP-HFPI initially improves dramatically and then stabilizes with minor variation. Based on this observation, we numerically confirm that setting $ \rho=0.5 $ can achieve good convergence performance in our simulation. Fig.$\, \ref{fig:epsilon} $ shows how the performance of FP-HFPI is affected by the inner loop tolerance $\epsilon_2$ in HFPI when the transmit SNR is 30 dB. As the $ \epsilon_2 $ increases from $ 10^{-4} $ to $ 10^0 $, the required average CPU time decreases by another order of magnitude, while the sum-rate does not suffer a significant loss (less than 0.2\%). It shows the tradeoff between performance and complexity for FP-HFPI.

\subsection{Performance Comparison under perfect CSIT}
\par
\begin{figure*}
	\begin{center}
		\begin{minipage}{0.3\textwidth}
			\includegraphics[width=1\linewidth]{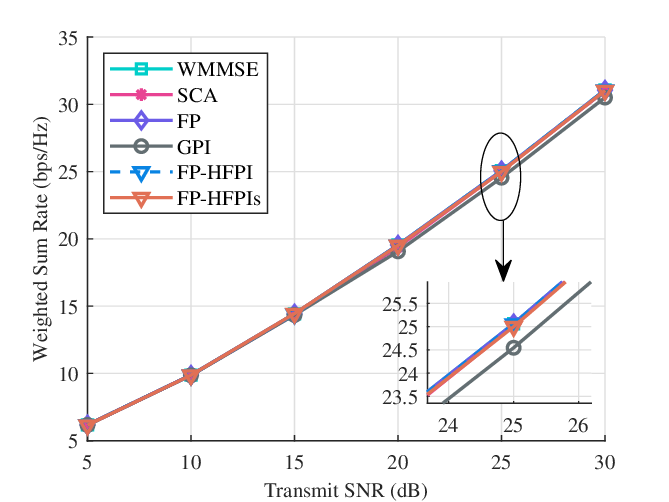}
			\caption{Weighted sum rate versus transmit SNR for different algorithms when $ L=4 $, $K=4 $.}
			\label{img:sum_rate_vs_SNR}
		\end{minipage}
	\begin{minipage}{0.3\textwidth}
				\includegraphics[width=1\linewidth]{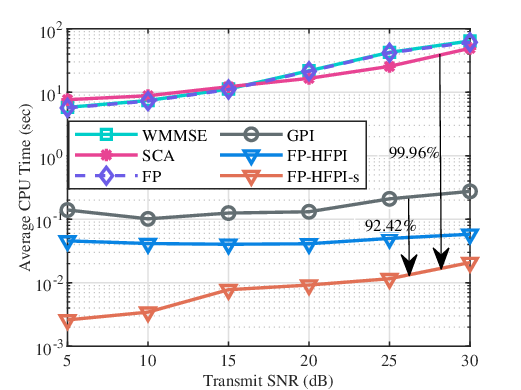}
		
		\caption{Averaged CPU time versus transmit SNR for different algorithms when $ L=4 $, $K=4 $.}
		\label{img:performance_vs_SNR}
	\end{minipage}
\begin{minipage}{0.3\textwidth}
			\includegraphics[width=1\linewidth]{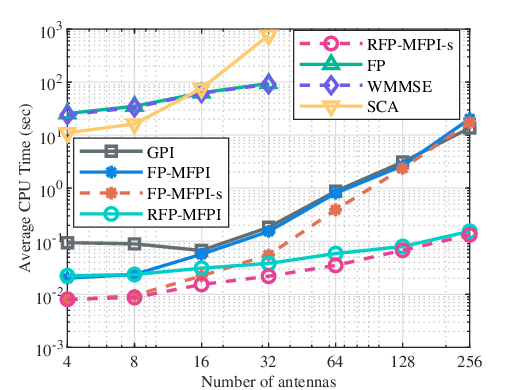}
	\caption{Averaged CPU time versus the number of transmit antennas when $K=4  $, SNR$=20  $dB.}
	\label{img:performance_vs_Nt}
\end{minipage}
	\end{center}
\end{figure*}

 Fig.$\, \ref{img:sum_rate_vs_SNR} $ shows that the proposed algorithms are effective to solve the original WSR problem without performance loss compared with other classical methods in all transmit SNR regions. Fig.$\, \ref{img:performance_vs_SNR} $ illustrates the corresponding average CPU time versus transmit SNR. In comparison to all existing methods, the average CPU time of the proposed algorithms are significantly reduced in all transmit SNR regimes. In particular, compared with SCA and GPI, the average CPU time of FP-HFPI-s reduces 99.96\% and 92.42\%, respectively.

%


\par Fig.$ \,\ref{img:performance_vs_Nt} $ illustrates the average CPU time for different algorithms versus the number of transmit antennas when the transmit SNR is $ 20 $ dB and the number of users is $ K=4 $. Despite the exponential increase in transmit antennas, the average CPU time of our proposed algorithms remains under 1\% of the time required by other algorithms that necessitate optimization solvers. Compared to the GPI method, we also have substantial CPU time reduction at the low antennas region, which comes from our stable convergence process. In contrast, GPI has to adjust the approximation parameters to guarantee the tight approximation. Furthermore, it has been shown that our proposed reduced FP-HFPI (RFP-HFPI) is much more computation efficient than all other methods when the number of transmit antennas increases exponentially. This greatly reduces the computational complexity of the transceiver design for RSMA and represents an important step forward in its further application in 6G. 

\begin{figure*}
	\begin{center}
		\begin{minipage}{0.3\textwidth}
			\includegraphics[width=1\linewidth]{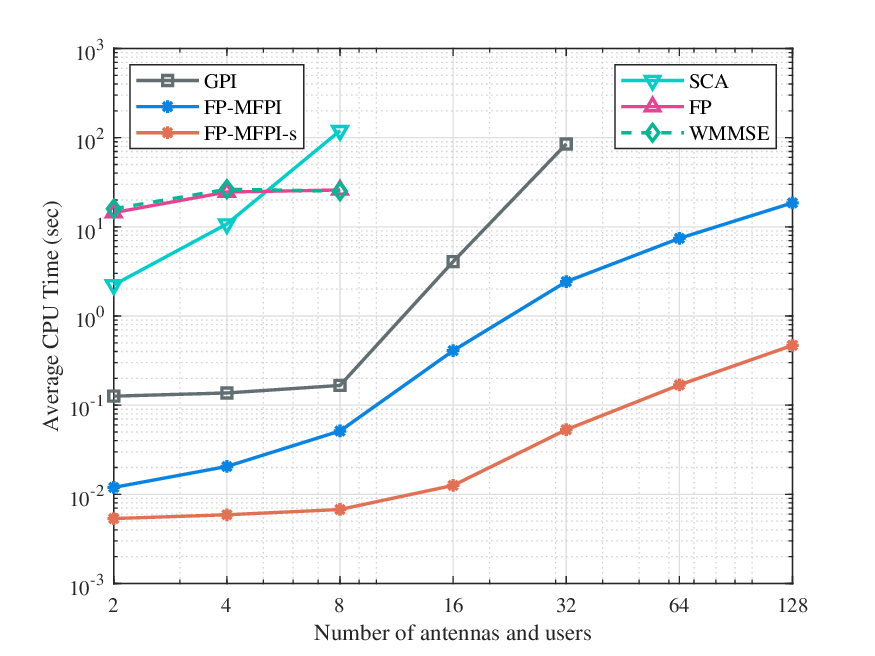}
		\caption{Averaged CPU time versus transmit antennas with equal users, i.e., $ L=K $ when SNR$=20  $dB.}
		\label{img:sum_rate_vs_Nt}
		\end{minipage}
		\begin{minipage}{0.3\textwidth}
			\includegraphics[width=1\linewidth]{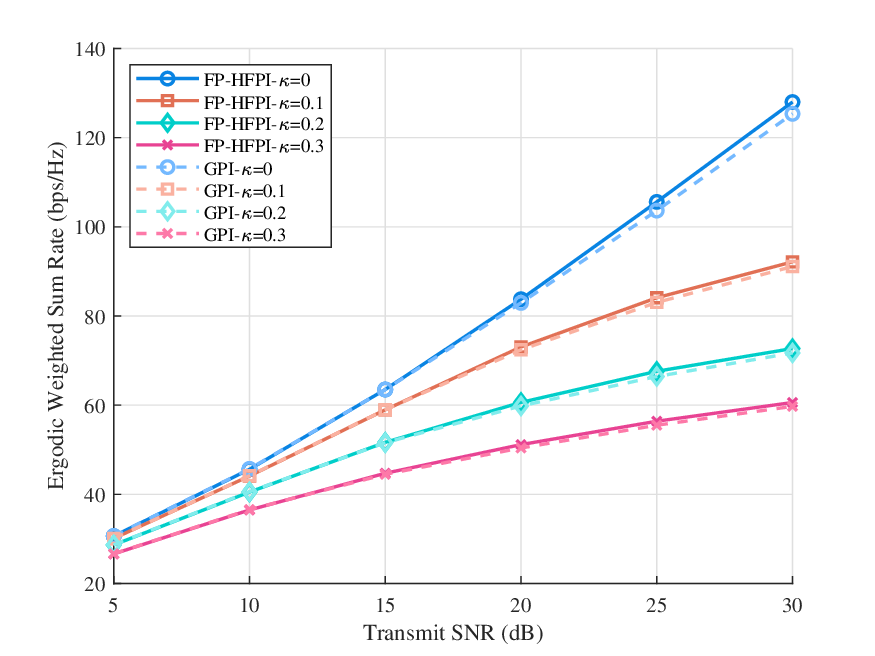}
			
			\caption{Ergodic WSR versus transmit SNR for different quality of CSIT when $ L=32 $, $K=16 $.}
			\label{img:erfodic SR_vs_SNR}
		\end{minipage}
		\begin{minipage}{0.3\textwidth}
			\includegraphics[width=1\linewidth]{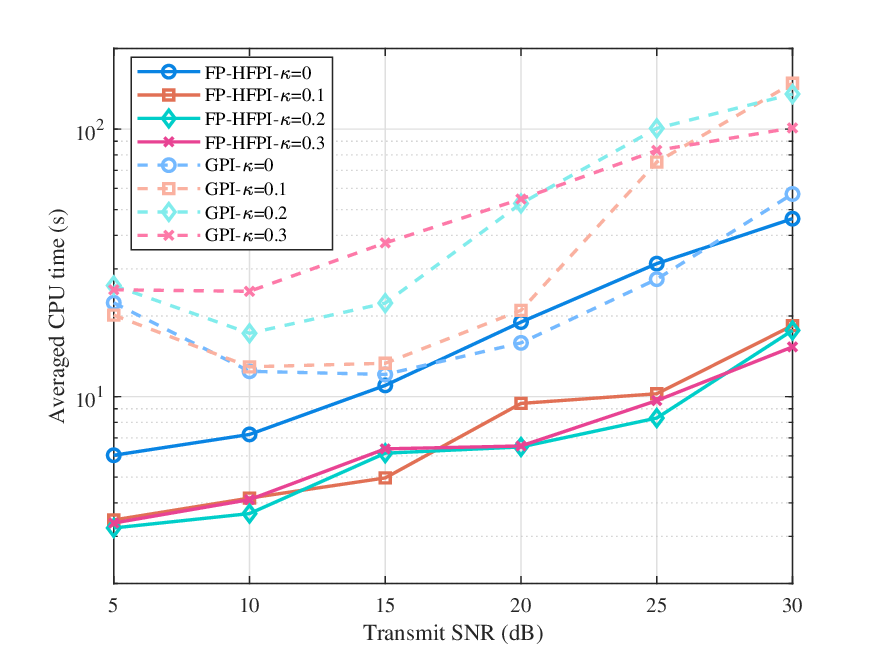}
			\caption{Averaged CPU time versus transmit SNR for different quality of CSIT when $L=32, K=16  $.}
			\label{img:cpu_vs_SNR_im}
		\end{minipage}
	\end{center}
\end{figure*}

\par Fig.$ \,\ref{img:sum_rate_vs_Nt} $ shows the average CPU time for different algorithms versus the number of transmit antennas with equal number of users, i.e., $ L=K $ when the transmit SNR is $ 20 $ dB. While both FP-HFPI and GPI have the same theoretical complexity order of $ \mathcal O(KL^3) $, it is clear that vectorizing a compact matrix into a high dimensional vector reduce the efficiency of the GPI algorithm. This is more obvious when the number of user $ K $ has the same magnitude with the number of transmit antennas $ L $ as shown in Fig.$\,  \ref{img:sum_rate_vs_Nt} $. GPI fails to address such a large-scale problem within reasonable time consumption, especially when the number of transmit antennas and users exceeds 32. In contrast, our proposed FP-HFPI remains efficient to address such large-scale problem. The comparable average CPU time for these two algorithms is observed only when $ L>>K  $, for example, with $ L\geq 16$ and $ K=4 $ as depicted in Fig.$ \, \ref{img:performance_vs_Nt} $.

\subsection{Performance Comparison under imperfect CSIT}

\par Fig.$ \,\ref{img:erfodic SR_vs_SNR} $ and Fig.$ \,\ref{img:cpu_vs_SNR_im} $ respectively show the performance of two CVX-free methods in terms of ergodic WSR (computed under the true channel) and averaged CPU time, with respect to the transmit SNR under the imperfect CSIT setup with various CSIT qualities. The number of transmit antenna is set to $L=32$ and the number of users is $K=16$.  In this large-scale setting, conventional optimization algorithms that rely on CVX, such as WMMSE, standard FP, and SCA are not applicable due to their extremely demonstrates that the proposed algorithm consistently achieves the same or even superior ergodic WSR compared to the GPI algorithm. Fig.$ \,\ref{img:cpu_vs_SNR_im} $ further shows that the computational CPU time of the proposed algorithm is lower than the CPU time required by GPI.
%
%
%

\section{Conclusion}\label{Sec:concu}
\par In this study, we present the optimal beamforming structure  and common rate allocation for 1-layer RSMA in a multiuser multi-antenna downlink transmission network. Our findings provide fundamental insights to the optimal beamforming design for common and private messages, revealing that they share similarities with conventional multicast and unicast messages, respectively, in the physical transmission perspective. Our results also reveal that the optimal beamforming structure is a weighted MMSE filter with a group channel direction for common stream and a single channel direction for private stream. We further demonstrate that the optimal beamforming solution has an inherent low-dimensional structure, allowing us to decrease the dimension of beamformers when the number of transmit antennas is extremely large. Based on the optimal beamforming structure, we further develop a numerical algorithm, namely FP-HFPI, to solve the WSR problem efficiently. Our algorithm provides near-optimal performance with low computational complexity compared to the state-of-the-art methods.

The discovered optimal beamforming structure is not limited to the WSR problem of 1-layer RSMA. It can also be applicable to other fundamental problems in 1-layer RSMA frameworks, such as power minimization, max-min fairness, and energy efficiency maximization problems. Additionally, it can also be extend to other RSMA frameworks, such as 2-layer hierarchical RS \cite{Dai2016} and the generalized RS \cite{mao2018}. Due to the generalization of RSMA over existing multiple access techniques, the proposed algorithms can also be applied for SDMA and NOMA. In addition, the developed optimization framework can be easily extended to MIMO BC with multiple antennas at the receivers or massive MIMO systems with imperfect SIC, which we plan to explore in our future work.
\begin{appendices}

	\section{Proof of Proposition \ref{pro: power constraint}}\label{Appendix: proof of power constraint}
The proof of Proposition \ref{pro: power constraint} is based on the complementary slackness conditions of the original problem \eqref{P1}. To be specific, we first introduce the following Lemma \ref{lemma: mu>0}.
	
 \begin{lemma}\label{lemma: mu>0}
 	For any non-zero stationary point $ \mathbf W^\diamond $ of problem (\ref{P1}), the corresponding Lagrangian dual variable $ \acute\mu^\diamond $ for the transmit power constraint \eqref{eq:power constraint} must be positive, i.e., $ \acute\mu^\diamond>0 $.
 \end{lemma}
	
	\textit{Proof:} Considering a non-zero stationary point $ \mathbf W^\diamond  $ of problem (\ref{P1}), there exists a Lagrangian dual variable $\acute\mu^\diamond  $ satisfying the KKT
	condition of problem \eqref{P1} as follows.
	\begin{subequations}\label{KKT for original}
		\begin{align}
\label{OPKKT:w0}		&	\sum_{k=1}^K \acute\lambda^\diamond _k \frac{\partial r_{0,k}}{\partial \mathbf w_0}-2\acute\mu^\diamond  \mathbf w^\diamond _0=0,\\
\label{OPKKT:wk}		&  \sum_{i=1}^K \delta_i \frac{\partial r_{i,i}}{\partial \mathbf w_k}+	\sum_{j=1}^K \acute\lambda_j^\diamond  \frac{\partial r_{0,j}}{\partial \mathbf w_k}-2\acute\mu^\diamond \mathbf w^\diamond _k=0, \forall k\in\mathcal K,\\
\label{OPKKTslack}	& \acute\mu^\diamond  \left(\mathrm{tr}(\mathbf w_0^\diamond \mathbf w_0^{\diamond^H})+\sum_{k=1}^K\mathrm{tr}(\mathbf w_k^\diamond\mathbf w_k^{\diamond^H}) - P_{t}\right)=0,\\
\label{OPKKT:mu}		& \acute\mu^\diamond\geq 0,
		\end{align}
	\end{subequations}
	where \eqref{OPKKT:w0} and \eqref{OPKKT:wk} are stationary conditions; \eqref{OPKKTslack} is the complementary slackness conditions; and \eqref{OPKKT:mu} is the dual feasibility condition. Note that here the dual variable $ \acute\mu $ is introduced for the original problem \eqref{P1}, which should not be confused with the dual variable $ \mu $ for subproblem \eqref{P5}.
	
	Next we prove Lemma \ref{lemma: mu>0} by contradiction. Specifically, we first assume that there exists a Lagrangian dual variable $ \acute\mu^\diamond=0 $ such the equations in  \eqref{KKT for original} hold, then we prove that under such assumption the KKT conditions in \eqref{KKT for original} cannot hold. We first rewrite the achievable rates of user $ k $ for decoding stream $ s_0 $ and $ s_k $ as
	\begin{equation*}
\begin{split}
			r_{0,k}
			\!&=\!\log\!\left(\sum_{j=0}^K |\mathbf h_k^H\mathbf w_j|^2\!+\!\sigma_k^2 \!\right)\!\!-\!\log\left( \sum_{j=1}^K |\mathbf h_k^H\mathbf w_j|^2+\sigma_k^2 \right),\\
			r_{k,k}\!&=\!\log\!\left(\sum_{j=1}^K |\mathbf h_k^H\mathbf w_j|^2\!+\!\sigma_k^2\! \right)\!\!-\!\log\left( \sum_{j=1,j\neq k}^K\!\!\! |\mathbf h_k^H\mathbf w_j|^2\!+\!\sigma_k^2\! \right).
\end{split}
	\end{equation*}
	Then, we calculate the partial derivatives of $ r_{0,k} $ with respect to $ \mathbf w_0 $, yielding
	\begin{equation}\label{partial z00k}
		\frac{\partial r_{0,k}}{\partial \mathbf w_0}=\mathbf h_k z_{0,0,k},
	\end{equation} 
	where
	\begin{equation}\label{z00k}
		z_{0,0,k}=2\left(\sum_{j=0}^K |\mathbf h_k^H\mathbf w_j|^2+\sigma_k^2 \right)^{-1}\mathbf h_k^H\mathbf w_0.
	\end{equation} 
Also, by taking the derivatives of $ r_{0,i} $ and $ r_{i,i} $ with respect to $ \mathbf w_k $, we have
\begin{subequations}\label{partial zkii}
	\begin{align}
		&\frac{\partial r_{0,i}}{\partial \mathbf w_k}=\mathbf h_i z_{k,0,i}, &
& \frac{\partial r_{i,i}}{\partial \mathbf w_k} =\mathbf h_i z_{k,i,i},
	\end{align}
\end{subequations}
where 
\begin{equation*}
	\begin{split}
			z_{k,0,i}\!&=\!2\!\left\{\!\!\left(\!\!\sum_{j=0}^K \!|\mathbf h^H_i\mathbf w_j|^2\!+\!\sigma_i^2 \!\!\right)^{\!\!\!-1}\!\!\!\!\!\!-\!\!\left(\!\sum_{j=1}^K |\mathbf h^H_i\mathbf w_j|^2+\sigma_i^2 \!\!\right)^{\!\!\!-1}\!\!\right\}\mathbf h_i^H\mathbf w_k,\\
			z_{k,i,i} &=2\left(\sum_{j=1}^K |\mathbf h^H_i\mathbf w_j|^2+\sigma_i^2 \right)^{-1}\mathbf h_i^H\mathbf w_k\nonumber\\
	&	\quad-2\left(\sum_{j=1,j\neq i}^K |\mathbf h_i^H\mathbf w_j|^2+\sigma_i^2 \right)^{-1}\!\!\!\!\!\mathbf h_i^H\mathbf w_k, \hfill i\neq k,
\end{split}
\end{equation*}

and	
\begin{equation*}\label{key}
		z_{k,k,k}=2\left(\sum_{j=1}^K |\mathbf h_i^H\mathbf w_j|^2+\sigma_i^2 \right)^{-1}\mathbf h_i^H\mathbf w_k
\end{equation*}

By substituting \eqref{partial z00k} and \eqref{z00k} into \eqref{OPKKT:w0}, and using the assumption $ \acute{\mu}^\diamond =0 $, we have
\begin{equation}\label{zero w0}
	\sum_{k=1}^K \acute{\lambda}_k^\diamond \mathbf h_k \left(\sum_{j=0}^K |\mathbf h_k^H\mathbf w_j|^2+\sigma_k^2 \right)^{-1}\mathbf h_k^H\mathbf w_0^\diamond=0.
\end{equation}
Note that equation \eqref{zero w0} holds if and only if $ \mathbf w_0^\diamond=\mathbf 0$, which results in $ z_{k,0,i}=0,\forall k,\forall i\in\mathcal K$. Then left-multiplying (\ref{OPKKT:wk}) by $ \mathbf w_k^{\diamond^H} $ yields
\begin{equation}\label{OPKKT partial w_k}
	\sum_{i=1}^K \delta_k\mathbf w_k^{\diamond^H}\frac{\partial r_{i,i}}{\partial \mathbf w_k}\bigg|_{\mathbf w_k=\mathbf w_k^\diamond}=0,\forall k\in\mathcal K,
\end{equation}
where we adopt $ z_{k,0,i}=0 $ and the assumption $ \acute{\mu}^\diamond =0 $. Summing \eqref{OPKKT partial w_k} over $ k=1,2,\cdots,K $ and rearranging the terms, we further obtain 
\begin{equation}\label{key}
	\sum_{i=1}^K\delta_i\frac{\sum_{k=1}^K|\mathbf h_i^H\mathbf w_k|^2}{\sum_{j=1}^K|\mathbf h_i^H\mathbf w_j|^2+\sigma_i^2}=\sum_{i=1}^K\delta_i\frac{\sum_{k=1,k\neq i}^K|\mathbf h_i^H\mathbf w_k|^2}{\sum_{j=1,j\neq i}^K|\mathbf h_i^H\mathbf w_j|^2+\sigma_i^2}.
\end{equation}
By further utilizing the equation $\frac{x}{1+x}=1-\frac{1}{1+x}  $, we have
\begin{equation}\label{eq: full}
	\sum_{i=1} ^K\!\!\delta_i\sigma_i^2 \!\!\left(\! \sum_{j=1}^K|\mathbf h_i^H\mathbf w_j|^2\!+\!\sigma_i^2\! \! \right)^{\!\!\!-1}\!\!\!\!\!\!=	\sum_{i=1}^K \!\!\delta_i\sigma_i^2 \!\!\left( \sum_{j=1,j\neq i}^K\!\!\!\!\!|\mathbf h_i^H\mathbf w_j|^2\!+\!\sigma_i^2 \!\! \right)^{\!\!\!-1}\!\!\!\!\!.
\end{equation}
Note that (\ref{eq: full}) holds if and only if $ \mathbf w_i^\diamond=\mathbf 0,\forall i\in\mathcal K $. Combining with $ \mathbf w_0^\diamond=\mathbf 0 $, we have $ \mathbf W^\diamond=\mathbf 0 $, which is inconsistent with the fact that $ \mathbf W^\diamond $ being a non-zero stationary point. Thus, for any non-zero stationary point $\mathbf W^\diamond$, the corresponding $\acute\mu^\diamond$ must be positive.$ \hfill \blacksquare $

Based on Lemma \ref{lemma: mu>0}, the complementary slackness condition given by (\ref{OPKKTslack}) and the equivalence of \eqref{P1} and \eqref{P3}, we directly obtain Proposition \ref{pro: power constraint}. $ \hfill \blacksquare $

\section{Derivation of The Update Equation \eqref{update_lambda}} \label{Appendix: proof of FP}
The complementary slackness conditions \eqref{kkt for lambda1} and \eqref{kkt for mu} can be rewritten as 
	\begin{equation}\label{key}
		\begin{split}
			\lambda_k^\star(h_{0,m}(\bm\lambda^\star,\mu^\star)-h_{0,k}(\bm\lambda^\star,\mu^\star)+\rho-\rho)&=0,\\
			\mu^\star(\mathrm{tr}(\mathbf W^\star\mathbf W^{\star^H})-P_t+\rho-\rho)&=0.
		\end{split}
	\end{equation}
	Further, moving the terms $ \lambda_k^\star(h_{0,k}(\bm\lambda^\star,\mu^\star)+\rho) $ and $ \mu^\star(P_t+\rho) $ to the right-hand side of the corresponding equations yields
	\begin{equation}\label{app2}
		\begin{split}
			\lambda_k^\star(h_{0,m}(\bm\lambda^\star,\mu^\star)+\rho)&=\lambda_k^\star(h_{0,k}(\bm\lambda^\star,\mu^\star)+\rho),\\
			\mu^\star(\mathrm{tr}(\mathbf W^\star\mathbf W^{\star^H})+\rho)&=\mu(P_t+\rho).
		\end{split}
	\end{equation} 	
	By dividing both sides of the equations in \eqref{app2} by the terms $ \lambda_k^\star(h_{0,k}(\bm\lambda^\star,\mu^\star)+\rho) $ and $ \mu^\star(P_t+\rho) $, respectively, we obtain
	\begin{equation}\label{fixed point}
		\begin{split}
			\lambda_k^\star&=\frac{h_{0,m}(\bm\lambda^\star,\mu^\star)+\rho}{h_{0,k}(\bm\lambda^\star,\mu^\star)+\rho}\lambda_k^\star\triangleq \phi_k(\bm\lambda^\star,\mu^\star),\\
			\mu^\star&=\frac{\mathrm{tr}(\mathbf W^\star\mathbf W^{\star^H})+\rho}{P_t+\rho}\mu^\star\triangleq \psi(\bm\lambda^\star,\mu^\star).
		\end{split}
	\end{equation}
	Equation \eqref{fixed point} reveals that the optimal dual variables $ \lambda_k^\star $ and $ \mu^\star $ must be the fixed points of iterative functions $ \phi_k(\bm\lambda,\mu) $ and $ \psi(\bm\lambda,\mu) $. It is intuitive to directly update the dual variables at iteration $ [t+1] $ with $ \lambda_k^{[t+1]}= \phi_k(\bm\lambda^{[t]},\mu^{[t]})$ and $ \mu^{[t+1]}=\psi(\bm\lambda^{[t]},\mu^{[t]}) $. However, the updated dual variables $ \bm\lambda^{[t+1]} $ does not satisfy the KKT condition \eqref{partial a} since $\phi_k(\bm\lambda^{[t]},\mu^{[t]})\leq \lambda_k^{[t]}  $. Therefore, we propose to project the point $ \bm\lambda^{[t+1]} $ back onto the hyperplane $ \mathcal{H}=\{\lambda_k\geq 0: \sum_{k=1}^{K} \lambda_k=\max_k\{\delta_k\}\}$ by further updating the dual variable $ \lambda_m $ with
	\begin{equation}\label{key}
		\lambda_m^{[t+1]}=\lambda_m^{[t]}\!+\!\sum_{k=1}^K\left(\lambda_k^{[t]}-\phi_k(\bm\lambda^{[t]},\mu^{[t]})\right).
	\end{equation}
	Finally, we end up with the proposed HFPI update equation \eqref{update_lambda}.

\begin{figure*}[htbp]
	\begin{equation}\label{partial lambda_k}
		\begin{aligned}
			\frac{\partial h_{0,k}(\bm\lambda,\mu)}{\partial \lambda_k}	&=2\sum_{i=1}^K\big||\beta_{k}|^2\mathbf h_k^H(\lambda_k\mathbf A_{k}+\mathbf B_{k})^{-1}\mathbf b_{i}\big|^2\mathbf h_k^H(\lambda_k\mathbf A_{k}+\mathbf B_{k})^{-1}\mathbf h_k\\
			&+2\big|\sqrt{1+\alpha_{0,k}}\beta_{0,k}-|\beta_{0,k}|^2\mathbf h_k^H(\lambda_k\mathbf A_{0,k}+\mathbf B_{0,k})^{-1}(\lambda_k \mathbf a_{0,k}+\mathbf b_{0,k})\big|^2\mathbf h_k^H(\lambda_k\mathbf A_{0,k}+\mathbf B_{0,k})^{-1}\mathbf h_k,\\
			\frac{\partial q(\bm\lambda,\mu)}{\partial \mu}	&= -2\sum_{k=1}^K \left((\mathbf H\mathbf \Theta_p\mathbf H^H+\mu\mathbf I)^{-1}d_{k,k}\mathbf h_k  \right)^H(\mathbf H\mathbf \Theta_p\mathbf H^H+\mu\mathbf I)^{-1} \left((\mathbf H\mathbf \Theta_p\mathbf H^H+\mu\mathbf I)^{-1}d_{k,k}\mathbf h_k  \right)\\
			&-2 \left((\mathbf H\mathbf \Theta_c\mathbf H^H+\mu\mathbf I)^{-1}\mathbf H  \mathbf d_c\right)^H(\mathbf H\mathbf \Theta_c\mathbf H^H+\mu\mathbf I)^{-1} \left((\mathbf H\mathbf \Theta_c\mathbf H^H+\mu\mathbf I)^{-1}\mathbf H  \mathbf d_c \right)
		\end{aligned}
	\end{equation} 
	\rule[-1pt]{\linewidth}{0.05em}
\end{figure*}
	\section{Proof of Proposition \ref{pro: only}} \label{Appendix:proof of unieq}
	We prove this proposition by introducing the following Lemma \ref{sss}.
	\begin{lemma}[\cite{boyd2004convex} ]\label{sss}
		For a convex problem, the optimal solution is unique if $ \mathcal L(\mathbf x,\bm\eta^\star) $ is strictly convex function of $ \mathbf x $, where $ \mathcal L $ refers to the corresponding Lagrangian function, $ \mathbf x $ is the primal variable and $ \bm\eta^\star $ is the optimal dual variable.
	\end{lemma}
	Recall that the full power consumption property stated in Corollary \ref{corollary equality} indicate that the power constraint must be satisfied with equality and thus $ \mu^\star>0  $ holds. Therefore, the Hessian matrices of Lagrangian function with respect to the beamforming vectors $ \mathbf w_0, \mathbf w_k,k\in\mathcal K $ are given by
	\begin{equation}\label{key}
		\begin{split}
			\nabla^2_{\mathbf w_0}\mathcal L(\mathbf W,y,\bm\lambda^\star,\mu^\star)&=\mathbf H\mathbf \Theta_c^\star\mathbf H^H+\mu^\star\mathbf I,\\
			\nabla^2_{\mathbf w_k}\mathcal L(\mathbf W,y,\bm\lambda^\star,\mu^\star)&=\mathbf H\mathbf \Theta_p^\star\mathbf H^H+\mu^\star\mathbf I,k\in\mathcal K.
		\end{split}
	\end{equation}    
	The positive definiteness of these Hessian matrices implies that $ L(\mathbf W,y,\bm\lambda^\star,\mu^\star) $ is a strictly convex function of $ \mathbf W $. Together with Lemma \ref{sss}, we conclude that there is a unique pair of $ \{\mathbf W^\star,y^\star, \bm\lambda^\star, \mu^\star\} $ satisfying the KKT conditions \eqref{kkt}, which also serves as the unique optimal solution of the convex problem \eqref{P5}.$ \hfill\blacksquare$

\section{Proof of Lemma \ref{lemma: monotone}}	\label{Appendix:monotic}
	This could be verified by the fact that $ \partial h_{0,k}(\bm\lambda,\mu)/\partial \lambda_k> 0 $ and $ \partial q(\bm\lambda,\mu)/\partial \mu<0 $. By the chain rule, the partial derivative of $ h_{0,k}(\bm\lambda,\mu) $ and $  q(\bm\lambda,\mu)$ with respect to $ \lambda_k $ and $ \mu $ are given by \eqref{partial lambda_k}, where 
	\begin{equation*}
		\begin{aligned}
			\mathbf A_{0,k}\!&=|\beta_{0,k}|^2\mathbf h_k\mathbf h_k^H, \!\!& 
			\mathbf A_{k}\!&=\mathbf A_{0,k},\\
			\mathbf B_{0,k}\!&=\!\!\!\!\!\sum_{j=1,j\neq k}^K\!\!\!\!\!\lambda_j|\beta_{0,j}|^2\mathbf h_j\mathbf h_j^H\!+\!\mu\mathbf I,\!\!&
			\mathbf B_{k}\!&=\mathbf B_{0,k}\!+\!\sum_{j=1}^K\delta_j|\beta_{j}|^2\mathbf h_j\mathbf h_j^H,\\
			\mathbf b_{0,k}\!&=\!\!\!\!\!\sum_{j=1,j\neq k}^K  \!\!\!\!\!\sqrt{1+\alpha_{0,j}}\beta_{0,j}\lambda_j\mathbf h_j,\!\!&
			\mathbf b_{i}\!&=\sqrt{1+\alpha_{i}}\beta_{i}\delta_i\mathbf h_i,\\
			\mathbf a_{0,k}\!&=\sqrt{1+\alpha_{0,k}}\beta_{0,k}\mathbf h_k.
		\end{aligned}
	\end{equation*}

	Note that we initialize Algorithm \ref{alg:proposed} with a non-zero initial point $ \mathbf W^{[0]} $, leading to the optimal auxiliary variable $ \beta_{i,k}^\star $ being non-zero as well. Therefore, we always have $ \partial h_{0,k}(\bm\lambda,\mu)/\partial \lambda_k> 0 $ and $  \partial q(\bm\lambda,\mu)/\partial \mu<0$.$ \hfill\blacksquare $

\end{appendices}
\bibliographystyle{IEEEtran}
\bibliography{reference}
%
%

\end{document}